\documentclass[11pt,a4paper]{amsart}
\usepackage[hypertex]{hyperref}
\usepackage{subfigure,caption2}
\usepackage{amssymb}

\usepackage{graphicx}
\usepackage[rflt]{floatflt}
\usepackage{pstcol}
\usepackage{pst-node}
\usepackage{pst-plot}
\numberwithin{equation}{section}
\newcommand{\smallpagebreak}{{\par\vspace{2 mm}\noindent}}

\newcommand{\demo}{\par\noindent{\it Proof.\/} \ }

\newcommand{\R}{{\mathbb R}}

\newcommand{\N}{{\mathbb N}}
\newcommand{\C}{{\mathbb C}}

\newcommand{\sign}{{\rm sign}\,}

\renewcommand{\Im}{{\rm Im}\,}
\renewcommand{\Re}{{\rm Re}\,}
\newcommand{\re}{{\rm Re}\,}
\newcommand{\im}{{\rm Im}\,}

\newcommand{\tr}{{\rm tr}\,}

\newcommand{\un}{{ \mathbb I}}
\theoremstyle{plain}
\newtheorem{Th}{Theorem}[section]
\newtheorem{Le}{Lemma}[section]
\newtheorem{Pro}{Proposition}[section]
\newtheorem{Cor}{Corollary}[section]
\theoremstyle{definition}

\newtheorem{Def}{Definition}[section]
\title{ Semi-classical  determination of exponentially small intermode
transitions for $1+1$ space-time scattering systems}
\author{Alain Joye} \author{Magali Marx}
\address[Alain Joye]{Institut Fourier, Unit\'e Mixte de Recherche CNRS-UJF 5582,
Universit\'e de Grenoble I, BP 74, F--38402 Saint Martin d'H\`eres Cedex, France}
\email{\href{mailto:}{alain.joye@ujf-grenoble.fr}}
\address[Magali Marx]{Institut Fourier, Unit\'e Mixte de Recherche CNRS-UJF 5582,
Universit\'e de Grenoble I, BP 74, F--38402 Saint Martin d'H\`eres Cedex, France}
\email{\href{mailto:}{magali.marx@ujf-grenoble.fr}}
\begin{document}
\begin{abstract}

We consider the semiclassical limit of systems of autonomous PDE's in
1+1 space-time dimensions in a scattering regime. We assume the matrix
valued coefficients are analytic in the space variable and we further suppose
that the corresponding dispersion relation admits real-valued modes only
with one-dimensional polarization subspaces. Hence a BKW-type analysis
of the solutions is possible. We typically consider time-dependent
solutions to the PDE which are carried asymptotically in the past and as
$x\rightarrow  -\infty$ along one mode only and determine
the piece of the solution that is carried for
$x\rightarrow +\infty$ along some other mode in the future.
Because of the assumed non-degeneracy
of the modes,
such transitions between modes are exponentially small in the
semiclassical parameter; this is an expression of the Landau-Zener mechanism.
We completely elucidate the space-time properties of the leading term
of this exponentially small wave, when the semiclassical parameter
is small, for large values of $x$ and $t$, when some avoided
crossing of finite width takes place between the involved modes.

                                %

                                %

%
\end{abstract}
\maketitle
\psset{unit=0.5em,linewidth=.05}
\section{Introduction}

Various physical models of wave propagation in space and time are
modelled by means of linear systems of autonomous Partial
Differential Equations (PDE's), with smooth or analytic
coefficients in the space variable $x\in\R^n$. The solutions to
such systems are usually difficult to compute in general and one
often resorts to asymptotic studies in the limit where the
wavelength involved is short with respect to the typical length
scale of the problem, on adapted time scales. This regime is often
called space-time adiabatic regime or semiclassical regime, due to
its relevance in Quantum Mechanics. Typical examples of that
situation are the short wavelength approximation of the wave
equation, of Maxwell equations and of the Klein-Gordon equation.
Similarly, the semiclassical analysis of the Dirac equation, of
the Schr\"odinger equation in solid state physics, or for
particles with spin in magnetic fields, and the Born-Oppenheimer
approximation in molecular physics belong to the same type of
problems. This is true also for certain Quantum systems whose
dynamics is constrained in nanotubes or waveguides. Also, the
study of shallow water waves in some linearized regime gives rise
to the linearized KdV or Boussinesq equations that share similar
properties. Plasma physics is another source of physically
relevant models entering this category.  See e.g. \cite{whitham},
\cite{bf},  \cite{hagedorn}, \cite{teufel}, \cite{bdt}...

From the mathematical point of view, it gives rise to singularly
perturbed problems for linear systems of PDE's. These problems are
tackled with success by means pseudo-differential operator
techniques and/or BKW methods which provide asymptotic solutions
up to errors of order $O(\varepsilon^m)$, where $\varepsilon$ is
the ratio of length scales, and $m$ depends on the peculiarities
of the problem. See the monographs \cite{hagedorn}, \cite{Fe:87},
\cite{Fe:89}, \cite{dimassi-sjostrand}, \cite{martinez},
\cite{teufel} for example.

The first step in the study of autonomous linear systems consists
is using separation of variables to reduce the problem to a
stationary system, parametrized by an energy variable conjugated
to the time variable. Superpositions of stationary solutions allow
to reconstruct solutions to the full time-dependent problem. Then,
one determines the dispersion relations or modes of the
corresponding symbol, and the associated polarization subspaces.
We will assume that all modes are real valued, i.e. we will
consider dispersive waves, according to \cite{whitham}. In the
semiclassical limit, when these real valued modes do not exhibit
crossings as the position and energy parameters vary, the dynamics
of the waves decouples inside the polarization subspaces in the
following sense: to leading order, independent waves driven by the
different scalar dispersion relations propagate along the
corresponding polarization subspaces without interacting. In
particular, transitions between isolated modes, or rather between
the corresponding polarization subspaces, are forbidden in the
semiclassical limit. Moreover, in the scattering limit, these
semiclassical transitions are typically of order
$O(\varepsilon^\infty)$, respectively $O(e^{-\Gamma/\varepsilon})$
for some $\Gamma >0$, in a smooth, respectively analytic context.
See e.g. \cite{hagedorn-joye}, \cite{nenciu-sordoni},
\cite{martinez-sordoni}, \cite{panati-spohn-teufel},
\cite{fermanian-gerard}, \cite{cdv1}, \cite{cdv2}, ... This
phenomenon goes under the name Landau-Zener mechanism, according
to the analysis of the adiabatic approximation of the time
dependent Schr\"odinger equation (in an  ODE context) which yields
transitions of this order between isolated eigenvalues, \cite{z},
\cite{l}, \cite{flynn-littlejohn}, \cite{joye94},
\cite{colindeverdiere},... Let us recall here that in case the
modes experience crossings at some point, the transitions  may be
of finite order in $\varepsilon$, indeed of zeroth order in some
cases, in the semiclassical limit \cite{hagedorn}. Their
determination is technically quite different and we do not address
these situations.

Although extremely small, the transitions between isolated modes
computed in the scattering limit are quite relevant from a
physical point of view in the various examples above. It is
therefore desirable for an ingoing wave prepared at large negative
times along one polarization mode, to determine the asymptotics as
$\varepsilon\rightarrow 0$ of the part of the waves that propagate
for large positive, but finite, times along another mode, be it a
transmitted or reflected wave. In a semiclassical context, to
achieve such a goal one is lead to further require the initial
wave to be well localized in energy. \smallpagebreak It is the aim
of this paper to determine such exponentially small transmitted
waves for quite general autonomous linear systems of PDE's in 1+1
space-time dimensions, when the coefficients are analytic and
possess limits they reach sufficiently fast as $|x|\rightarrow
\infty$.

\smallpagebreak

While the conditions allowing the determination of exponentially
small transitions between isolated modes for a variety of physical
situations are rather well understood now in a ODE context, or in
the language and setting sketched above, for stationary solutions,
see \cite{joye-kunz-pfister}, \cite{joye-pfister93},
\cite{joye94}, \cite{martin-nenciu}, \cite{ramond}, \cite{grigis},
\cite{joye-pfister95}, \cite{joye97}, \cite{fedotov-klopp},
\cite{fedotov-klopp2} \cite{hagedorn-joye04},
\cite{betz-teufel-1}, \cite{marx}, ... , it is well known that the
description of inter-mode transitions in a time-dependent context
requires more work. The only mathematical results we are aware of
regarding this issue concern the Born-Oppenheimer approximation in
molecular physics \cite{JH:04}.

The paper \cite{JH:04} is mainly motivated by molecular physics
considerations and the asymptotic descriptions provided there rely
heavily on peculiarities of the Born-Oppenheimer approximation
considered. However, as will become clear, the general strategy of
the analysis is actually model independent and, at the price of
sometimes substantial modifications, it can be adapted to fit the
various models and situations mentioned above. The importance and
frequency of the mechanism of inter-mode transitions in various
fields of applied mathematics is the main motivation for the
present work. Our aim is to extract practical conditions on a
system of PDE's in 1+1 space-time dimensions under which the
exponentially small pieces of propagating waves describing
inter-mode transitions in a scattering regime can actually be
computed, in the semiclassical limit. In that sense, the present
paper can be viewed as a generalization of \cite{JH:04}.

\vspace{.5cm}

Let us describe more precisely the autonomous systems we will be
dealing with, the type of results we get and the underlying
strategy we use to prove these results. Since it requires a fair
amount of notations and hypotheses to give a precise statement of
our main result,
Theorem \ref{th:astrans}, we remain at a rather informal level in this introduction.\\

Let $\mathcal{R}(x,
i\varepsilon\partial_{t},i\varepsilon\partial_{x})$ be the
differential operator:
\begin{equation}
\label{eq:diff_op}
\mathcal{R}(x,i\varepsilon\partial_{t},i\varepsilon\partial_{x})
=\sum\limits_{l\in\{0,\dots, m\},n\in\{0,\dots,
r\}}A_{ln}(x)(i\varepsilon\partial_{x})^{l}(i\varepsilon\partial_{t})^{n},
\end{equation}
where the $d \times d$ matrix valued coefficients $A_{ln}(x)$ are
independent of $t$ and analytic in $x$ a neighborhood of the real
axis. Assuming the matrices $A_{ln}(x)$ possess limit as
$|x|\rightarrow \infty$ which they reach fast enough, we want to
describe the small $\varepsilon$ behavior of certain solutions
$\phi(x,t,\varepsilon)$ to the evolution equation
\begin{equation}\label{evol}
\mathcal{R}(x,i\varepsilon\partial_{t},i\varepsilon\partial_{x})
\phi(x,t,\varepsilon)=0,
\end{equation}
for $x\in\R$, in the scattering regime $t$ large, in $L^2(\R)$.\\

The $d\times d$ matrix valued symbol, $R(x,E,k)$ corresponding to
$\mathcal{R}(x,
i\varepsilon\partial_{t},i\varepsilon\partial_{x})$ writes
\begin{equation}
R(x,E,k)=\sum\limits_{l\in\{0,\dots, m\},n\in\{0,\dots,
r\}}A_{ln}(x)k^{l}E^{n},
\end{equation}
where we call the dual variables $E$ and $k$ the energy and the
momentum variables respectively. The energy parameter will be
taken in a window $\Delta \subset \R$ specified below. The
associated dispersion relations or modes are defined as the set of
roots $\{k_j(x,E)\}$ of the polynomial equation in $k$, of degree
$md$, for $x\in\R$ and $E\in\Delta$,
\begin{equation}\label{det}
\mbox{det } R(x,E,k)=0.
\end{equation}
Our main assumption regarding the type of PDE we consider reads as
follows: we suppose there exists an energy window $\Delta$ such
that for all $E\in\Delta$, and all $x\in \R$, there exist $md$
distinct real valued modes $\{k_j(x,E)\}_{0\leq j\leq md}$. The
associated kernels $R(x,E, k_j(x,E))$, $j=1, \cdots, md$ are then
shown to be one-dimensional and their elements, denoted by
$\varphi_j(x,E)$, are the polarization vectors.

For comparison and illustration purposes, the case considered in
\cite{JH:04} corresponds to
$(i\varepsilon\partial_t+\varepsilon^2\partial_x^2/2+A_{0 0}(x))
\phi(x,t,\varepsilon)=0,$ where $A_{0 0}(x)=-V(x)$, is the
``electronic hamiltonian'', {\em i.e.} a $d\times d$ self-adjoint
matrix. It is assumed that $V(x)$ has non-degenerate eigenvalues
$\{e_1(x),\cdots,  e_d(x)\}$ with associated eigenvectors
$\{\varphi_1(x),\cdots,  \varphi_d(x)\}$. For large enough
energies $E$, (\ref{det}) yields $\mbox{det }(E-k^2/2 -V(x))=0$,
which provides the real valued modes $\{-\sqrt{2(E-e_1(x))},
\cdots, -\sqrt{2(E-e_d(x))},
  \sqrt{2(E-e_d(x))}, \cdots, \sqrt{2(E-e_d(x))}\}$ and
corresponding polarization vectors $\{\varphi_1(x),\cdots,
\varphi_d(x), \varphi_d(x), \cdots, \varphi_1(x)\}$.
\\

Our assumption is very close to the definition of linear
dispersive systems in nonuniform autonomous medium given in
\cite{whitham}, chapter 11. Such linear systems are characterized
there by the fact that the dispersion relation can be solved  in
the form of real roots $$E=W(k,x), \ \ \ \mbox{with}\ \ \
\partial_k^2W(k,x)\neq 0,$$ for $k$ real and $x\in\mathbb R$. This
notion is also reminiscent of the strictly hyperbolic equations
\cite{Tr}. In \cite{Tr}, a 1+1 first order partial differential
equation is called strictly hyperbolic in $x$ if it can be written
as:
$$\partial_{x}\Phi-A(x,t)\partial_{t}\Phi-B(x,t)\Phi=0,$$
where the matrix $A(x_{0},t_{0})$ has real and distinct
eigenvalues. If $A$ and $B$ only depend on $x$, these equations
are of the same type as \eqref{evol} for $r=m=1$. However, our
assumption and this notion are different, in general. The author
of  \cite{Tr} gives a characterization for strictly hyperbolic
systems of the form
$$\partial_{x}^{m}\Phi=\sum\limits_{l<m,l+p\leq
m}A_{lp}(x,t)\partial_{x}^{l}\partial_{t}^{p}\Phi$$ in terms of
the principal symbol. By contrast, our assumption concerns the
total symbol.
\\

Separation of variables allows  to construct solutions to
(\ref{evol}) by means of the formula
\begin{equation}\label{superp}
\phi(x,t,\varepsilon):=\int_\Delta
Q(E,\varepsilon)e^{-itE/\varepsilon} \psi_\varepsilon(x,E)\ dE,
\end{equation} where $\psi_\varepsilon(x,E)$ is a solution to the
energy dependent stationary problem
\begin{equation}\label{stat}
\widehat R(x,E,i\varepsilon\partial_x)\psi_\varepsilon(x,E)=0,
\end{equation}
with
\begin{equation}
\widehat R(x,E,i\varepsilon\partial_x)=\sum\limits_{l\in\{0,\dots,
m\},n\in\{0,\dots,r\}}A_{ln}(x)E^n (i\varepsilon\partial_{x})^{l},
\end{equation}
and the function $Q(\cdot ,\varepsilon):\Delta \mapsto \mathbb C$
is an energy density which ensures that $E$ belongs to the
prescribed window $\Delta$. The dependence of
$Q(\cdot,\varepsilon)$ on the parameter $\varepsilon$ will be used
to localize in energy the waves we want
to describe. \\

Equation (\ref{stat}) is a singularly perturbed system of ODE, to
which we apply complex BKW techniques. Making use of suitably
normalized polarization vectors $\varphi_j(x,E)$, we show that the
solutions of (\ref{stat}) can be expanded as
\begin{equation}
\psi_\varepsilon(x,E)=\sum_{j=1}^{md}c_j(x,E,\varepsilon)
e^{-i\int_0^x k_j(y,E) dy/\varepsilon}\varphi_j(x,E),
\end{equation}
where the $\mathbb C$ valued coefficients $c_j(\cdot,E,\varepsilon
)$ satisfy some linear ODE, which we analyze in the semiclassical
limit $\varepsilon\rightarrow 0$. The assumption $E\in\Delta$
imply that the factors $e^{-i\int_0^x k_j(y,E) dy/\varepsilon}$
are phases for all $x\in \mathbb R$ with distinct $k_j(x,E)$, and
the coefficients $c_j$ are constant in the semi-classical limit,
see e.g. \cite{joye-pfister95}, \cite{joye97},
\begin{equation}\label{1/2}
c_j(x,E,\varepsilon)=c_j(0, E,\varepsilon)+O(\varepsilon), \ \ \
j=1,\cdots, md.
\end{equation}
The hypotheses on the matrices $A_{ln}$ at infinity ensure the
existence of the limits $\varphi_j(\pm\infty,E)$, $k_j(\pm\infty,
E)$, $c_j(\pm\infty,E,\varepsilon)$, and the error term in
(\ref{1/2}) is uniform in $x$. In particular, the stationary
on-shell scattering process characterized by the $S$-matrix
$$S(E,\varepsilon)c(-\infty,E,\varepsilon)=c(+\infty, E,\varepsilon), \ \ \ \mbox{where}
\ \ \ c(+\infty, E,\varepsilon)=\left(\begin{matrix}c_1(+\infty,
E,\varepsilon) \cr c_2(+\infty, E,\varepsilon) \cr \vdots \cr
c_{md}(+\infty, E,\varepsilon)
\end{matrix}\right) $$
is well defined. Actually, in our analytic framework, the
off-diagonal elements of $S(E,\varepsilon)$ are exponentially
small, see below. Thus, for $|x|$ large enough, the solutions
(\ref{superp}) of the time dependent equation (\ref{evol}) behave
as
\begin{equation}\label{assol}
\phi(x,t,\varepsilon)\simeq \sum_{j=1}^{md} \int_\Delta
Q(E,\varepsilon) c_j(\pm\infty, E, \varepsilon) e^{-i(tE+x
k_j(\pm\infty,E))/\varepsilon}\varphi_j(\pm\infty,E)\ dE.
\end{equation}
Assuming the asymptotic dispersion relations $E\mapsto
k_j(\pm\infty,E)$ are invertible on $\Delta$, the asymptotic
solutions (\ref{assol}) are given by linear combinations of wave
packets associated with each mode and corresponding polarization.
The property (\ref{1/2}) shows that transitions between modes
induced by the evolution are
vanishing in the semi-classical limit.\\

We determine the asymptotics of certain exponentially small
transitions between modes for solutions that allow to define a
scattering process for $|x|$ large, in a time-dependent set up.
{\em Bona fide} scattering processes require the energy and the
modes we are interested to be such that there exists a mode
supporting ingoing waves on which we start our solution at time
$-\infty$ and that there exists  another  mode describing outgoing
waves at time $+\infty$ to which transitions are possible. There
exist systems of PDE's that intertain outgoing solutions
or ingoing solutions only. Our results do not provide interesting informations for such systems.\\

For definiteness, let us assume in that introduction, that the
energy of the waves is well localized around
$E_0\in\Delta\setminus\partial\Delta$ and that, for all
$x\in\mathbb R$, $\partial_{E}k_j(x, E_0)<0$. Our sign conventions
imply that the asymptotic group velocities are then positive, see
in particular Proposition \ref{stph}. This implies that such waves
travel from the left to the right and are polarized along
$\varphi_j(-\infty, E_0)$ in the remote past. Let us further
assume the mode $k_n$ supports outgoing solutions from the left to
the right as well, for $x\simeq +\infty$. The incoming waves are
thus characterized for $x$ large and negative by stationary
solutions corresponding to $c_k(-\infty, E,
\varepsilon)=\delta_{kj}$. Hence, the summand with label $n$ of
(\ref{assol}) corresponding to the coefficient $c_n(+\infty, E,
\varepsilon)$, $n\neq j$, determines the exponentially small piece
of the wave for $x$ in a neighborhood of $+\infty$ that has made
the transition from mode $k_j$ to mode $k_n$ in the course of the
evolution, for
times $t$ large and positive. \\

In order to compute the exponentially small asymptotics of the
coefficient  $c_n(+\infty, E, \varepsilon)$, one uses BKW
techniques. That is one considers the equation satisfied by these
coefficients in the complex plane and makes use of their
multivaluedness around points of degeneracy of the analytic
continuations of certain modes. As is well known, the complex BKW
method requires the existence of dissipative or canonical domains,
e.g. \cite{Fe:87}, \cite{Fe:89}, \cite{grigis},
\cite{joye-kunz-pfister}, \cite{joye97},
\cite{fedotov-klopp},\cite{fedotov-klopp2}, \cite{marx}, ... which
is not easy to prove. In our setup, we rely on the analysis of
\cite{joye97} which proves that in some avoided crossing regime,
dissipative domains exist. The notion of avoided crossing requires
the introduction of another parameter but we don't want to be
specific about this regime yet. Let us only mention here that
dissipative domains exist in particular when the mode $k_j(\cdot,
E)$ becomes almost degenerate with $k_l(\cdot, E)$, with either
$l=j-1$ or $l=l+1$, at one point only on the real axis. The
outcome of the analysis is the asymptotic formula for
$\varepsilon\rightarrow 0$
\begin{equation}\label{lz}
c_n(+\infty, E, \varepsilon)= \tau(E)e^{iS(E)/\varepsilon}
(1+O(\varepsilon))
\end{equation}
with $S(E)=\kappa(E)+ i\gamma(E)$, $\gamma(E)>0$ and $\tau(E)\in
\mathbb C^*$. The exponent $S(E)$ is given by some action integral
in the complex plane around the relevant complex degeneracy point
of the modes $k_j(\cdot, E)$ and $k_n(\cdot, E)$, see
(\ref{eq:wkb}), and the prefactor $\tau(E)$ possesses some
geometric meaning \cite{joye-kunz-pfister}. We localize our wave
packets in energy by considering typically Gaussian energy
densities of the form
\begin{equation}
Q(E,\varepsilon)=P(E,\varepsilon)e^{-(E-E_0)^2g/(2\varepsilon)},
\end{equation}
where $P$ has support in $\Delta$ and diverges at worst like a
polynomial
in $1/\varepsilon$ as $\varepsilon\rightarrow 0$. \\

With these ingredients, we prove in Theorem \ref{th:astrans} that
for $t>0$ large enough, and in the $L^2(\mathbb R)$ norm, the
piece of the wave function that has made the transition from the
mode $k_j$ to $k_n$, is given in the limit $\varepsilon\rightarrow
0$ by
\begin{eqnarray}\label{aswa}
\phi_{nj}(x,t,\epsilon)&\simeq& e^{-\alpha_*/\varepsilon}
N_*(\varepsilon)\int_{ k_n(+\infty,
\Delta)}e^{-\lambda_2(k-k_*)^2/(2\varepsilon)}
e^{-i(tE_n^+(k)+kx+\lambda_1(k-k^*))/\varepsilon}\ dk \nonumber\\
& &\qquad \qquad \qquad \qquad \qquad \qquad \qquad+
O(1/|t|^\beta) + o({\varepsilon^{3/4}} e^{-\Re
\alpha_*/\varepsilon}N_*(\varepsilon)).
\end{eqnarray}
Here $0<\beta<1/2$, $k \mapsto E_n^+(k)$ is the inverse function
of the asymptotic dispersion relation $E\mapsto k_n(+\infty, E)$.
The exponent $\alpha_*$, the average momentum $k_*$ and factors
$\lambda_1$, $\lambda_2$ (such that $\Re \lambda_2>0$) are
determined by the action integral $S$ and the energy density $Q$,
and the prefactor $N_*(\varepsilon)$ is polynomial at worse in
$1/\varepsilon$. The leading term in (\ref{aswa}) is of positive
$L^2$ norm, constant in time, and of order ${\varepsilon^{3/4}}$,
up to the prefactors. Moreover, as $\varepsilon\rightarrow 0$ and
$|t|\rightarrow \infty$ this wave is essentially carried on a ball
centered at $x=-\partial_k E_n^+(k_*)t$ of radius $\sqrt{t}$, in
the $L^2$ sense, see Proposition \ref{stph}. Finally, the error
terms are uniform in $\varepsilon$ and $t$, respectively.

The function (\ref{aswa}) corresponds to an exponentially small
free wave propagating according to the dispersion relation
$E_n^+(k)$ with Gaussian momentum profile (within the momentum
window $ k_n(+\infty, \Delta)$) centered around $k_*$. Note that
the error terms are negligible only for large enough times,
actually exponentially large times $t\simeq  e^{c/\varepsilon}$,
$c>0$. Let us emphasize one point revealed by the present analysis
and that of \cite{JH:04}. The average momentum $k_*$ does not
coincide with the naive guess $k_0\simeq k_n(+\infty, E_0)$, which
corresponds to energy conservation. It is actually dependent on
the choice of energy density $Q$. Similarly, the exponential decay
rate  $\alpha_*$ is not determined by the function $\gamma=\Im S$
only, but depends explicitly on the density $Q$ as well.

In other words, the piece of the wave function that has made the
transition is asymptotically given for small $\varepsilon$ and
large times by the solution to the linear evolution equation, in
(rescaled) Fourier space,
\begin{equation}\label{asedp}
i\varepsilon\partial_t f(t,k)=E_n^+(k)f(t,k), \ \ f(0,k)=
e^{-\alpha_*/\varepsilon}
N_*(\varepsilon)e^{-\lambda_2(k-k_*)^2/(2\varepsilon)}e^{-i\lambda_1(k-k_*)/\varepsilon}
{\chi}_{k_n(+\infty,\Delta)}.
\end{equation}
Finally, we mention that in case $E_n^+(k)$ is quadratic in $k$,
we can further compute the leading term explicitly, as in
\cite{JH:04}, which yields a freely propagating Gaussian, see
Lemma \ref{le:quad}. Also, our analysis applies to the description
of exponentially small reflected waves, as will be explained
below.

\vspace{.3cm}

The rest of the paper is organized as follows. The precise
hypotheses on the operator  $\mathcal{R}(x,
i\varepsilon_{t},i\varepsilon\partial_{x})$, are spelled out in
the next Section. Section \ref{sec:geneigen} is devoted to the
analysis of the corresponding stationary solutions. The BKW method
and the avoided crossing situation are presented in Section
\ref{sec:wkb}. The construction of time-dependent solutions to the
original problem and their scattering properties are given in
Section \ref{sec:t_depdt}. The precise semiclassical analysis in
the scattering regime of the time dependent asymptotic waves
describing inter-mode transitions is provided in Section
\ref{sec:transas}. Further properties of the asymptotic waves
are given in Section \ref{spti}. A technical Section closes the paper.

\section{Hypotheses for the differential operator}
\label{sec:hyp} We consider a differential operator defined by
\eqref{eq:diff_op} where a supplementary small parameter $\delta$
is included to define the avoided crossing regime in which the gap
between certain modes are small:
\begin{equation}\label{eq:diff_op2}
\mathcal{R}(x,i\varepsilon\partial_{t},i\varepsilon\partial_{x},\delta)
=\sum\limits_{l\in\{0,\dots, m\},n\in\{0,\dots,
r\}}A_{ln}(x,\delta)(i\varepsilon\partial_{x})^{l}(i\varepsilon\partial_{t})^{n}.
\end{equation}
We recall that $$\forall(l,n)\in\{0,\dots, m\}\times\{0,\dots,
r\},\quad \forall x\in\R,\quad \forall\delta\in[0,d_{0}],\quad
A_{ln}(x,\delta)\in\mathcal{M}_{d}(\C)$$ and we define:
$$R(x,E,k,\delta)=\sum\limits_{l\in\{0,\dots, m\},n\in\{0,\dots,
r\}}A_{ln}(x,\delta)k^{l}E^{n}.$$ Now, we describe the hypotheses
on the differential operator $\mathcal{R}$.
\begin{description}
\item[(H1)] There exist $Y>0$ and $d_{0}>0$ such that for any
$\delta\in[0,d_{0}]$ the matrix valued functions $z \mapsto
A_{ln}(z,\delta)$ are analytic in a strip $\rho_{Y}=\{z\in\C;\
|\im z|<Y\}$ and \\ $(z,\delta)\mapsto A_{ln}(z,\delta)$ is
$C^{3}$ for any $(z,\delta)\in\rho_{Y}\times[0,d_{0}]$.
\item[(H2)] There exist $\nu>1/2$, $c>0$ and $2 mr$ matrix valued $C^{2}$
functions $\{\delta\mapsto
A_{ln}(\pm\infty,\delta)\}_{l\in\{0,\dots, m\}}^{n\in\{0,\dots,
r\}}$ such that $\forall \delta\in[0,d_{0}]$,:
$$\sup\limits_{z\in\rho_{Y},\re
z<0}|\re
z|^{2+\nu}\|A_{ln}(z,\delta)-A_{ln}(+\infty,\delta)\|+\sup\limits_{z\in\rho_{Y},\re
z>0}|\re
z|^{2+\nu}\|A_{ln}(z,\delta)-A_{ln}(-\infty,\delta)\|<c.$$ Now, we
describe the assumption of avoided crossing. We assume that
$\Delta\subset\R$ is a compact interval with non-empty interior
such that, for any $E\in\Delta$:
\item[(H3)] For any $x\in\R$ and any $\delta\in[0,d_{0}]$, there are
$md$ real values
$\{k_{1}(x,E,\delta),k_{2}(x,E,\delta),\dots,k_{md}(x,E,\delta)\}$
such that $\det R(x,E,k,\delta)=0$.\smallpagebreak For any
$\delta\in[0,d_{0}]$, the values $k_{j}(x,E,\delta)$ have $md$
distinct limits as $x\rightarrow-\infty$ and as
$x\rightarrow+\infty$ that we denote by $k_{j}(\pm
\infty,E,\delta)$. The labels are chosen as follows.
\smallpagebreak When $\delta>0$, the functions $k_{j}(x,E,\delta)$
are distinct for $x\in[-\infty,+\infty]$ and are labelled by:
$$k_{1}(x,E,\delta)<k_{2}(x,E,\delta)<\dots<k_{md}(x,E,\delta).$$

When $\delta=0$, the functions $k_{j}(x,E,0)$ such that $\det
R(x,E,k,\delta)=0$ are given by $md$ real functions that have
$p(E)>0$ finitely many real crossings at
$x_{1}(E)<\dots<x_{p(E)}(E)$. Precisely, we assume that for some
fixed positive $\tilde{Y}$ and for any fixed $E\in\Delta$ that
\begin{itemize}
\item The functions $k_{j}(x,E,0)$ are labelled according to:
$$k_{1}(-\infty,E,0)<k_{2}(-\infty,E,0)<\dots<k_{md}(-\infty,E,0).$$
\item For any $j\in \{1,\dots,md\}$, the function $(z,E)\mapsto
k_{j}(z,E,0)$ is continuous on $\rho_{\tilde{Y}}\times\Delta$.
\item For any $j \in \{1,\dots,md\}$, the function $z\mapsto k_{j}(z,E,0)$ is analytic on
$\rho_{\tilde{Y}}$.
\item For any $l\in\{1,\dots,p(E)\}$, there exist exactly two
integers $(i,j)\in\{1,\dots,md\}^{2}$ such that:
$$k_{i}(x_{l}(E),E,0)=k_{j}(x_{l}(E),E,0).$$
Besides, we assume that
$$\partial_{x}(k_{i}-k_{j})(x_{l}(E),E,0)\neq 0.$$
\end{itemize}
For certain results, we also impose the condition that these
avoided crossings be generic in the sense of
\cite{oldHag,joye94,JH:04}.
\item[(H4)] Fix $E_{0}\in\Delta$. Near an avoided crossing $(x_{0}(E_{0}),E_{0})$ of $k_{i}$ and $k_{j}$, there
exist three functions $E\mapsto a(E)$, $E\mapsto b(E)$, $E\mapsto
c(E)$ such that, in a neighborhood of $E_{0}$,
\begin{enumerate}
\item The difference $k_{j}-k_{i}$ satisfies:
$$[k_{j}(z,E,\delta)-k_{i}(z,E,\delta)]^{2}=a^{2}(E)(z-x_{0}(E))^{2}+2c(E)(z-x_{0}(E))\delta+b^{2}(E)\delta^{2}+R_{3}(z-x_{0}(E),\delta),$$
where $R_{3}$ is a remainder of order $3$.
\item We have:
$$a(E)>0,\qquad b(E)>0,\qquad a^{2}(E)b^{2}(E)-c^{2}(E)>0.$$
\end{enumerate}
\end{description}
According to \cite{Ka}, we know a priori that the functions
$k_{j}$ are analytic in both variables except at the crossing
points. The assumptions $({\bf H{1}})$, $({\bf H{2}})$ and $({\bf
H{3}})$ imply analyticity in both variables at the real crossing
points:
\begin{Le}
\label{le:ana} Assume that $({\bf H{1}})$, $({\bf H{2}})$ and
$({\bf H{3}})$ are satisfied. Then, for $\Delta$ small enough,
\begin{enumerate}
\item the number $p(E)$ does not depend on $E\in\Delta$.
\item there exists $Y>0$ such that $(z,E)\mapsto k_{j}(z,E,0)$ is
analytic on $\rho_{Y}\times \Delta$, for any $j\in\{1,\dots,md\}$.
\item for $l\in\{1,\cdots, p(E)\}$, the function $E\mapsto x_{l}(E)$
is analytic on $\Delta$.
\end{enumerate}
\end{Le}
We will prove Lemma \ref{le:ana} in Section \ref{sec:techn}.
\smallpagebreak Similarly, assumptions ${\bf (H{1})}$ to ${\bf
(H4)}$ imply the following result:
\begin{Le}
\label{le:dev} Under assumptions ${\bf (H{1})}$ to ${\bf (H4)}$,
the functions $a$, $c$ and $b^{2}$ are analytic in a neighborhood
of $E_{0}$. Besides,
$a(E)=|\partial_{z}(k_{i}-k_{j})(x_{0}(E),E)|$.
\end{Le}
Lemma \ref{le:dev} is proven in Section \ref{sec:techn}.\\

Let us end this Section by noting here that our hypotheses imply
that the modes are real, but they do not guarantee that the $L^2$
norm is conserved under the time evolution. This question is
addressed in Section \ref{sec:t_depdt}.

\section{Generalized Eigenvectors}
\label{sec:geneigen} In this Section, we assume that $R$ and
$\Delta$ satisfy $\mathbf{(H3)}$, and we investigate the
properties of the modes, their corresponding polarization vectors
and the stationary solutions. For the time being, the parameter
$\delta>0$ is fixed and we drop it in the notation. The
generalized eigenvectors $\psi_{\varepsilon}(x,E)\in\C^{d}$ are
defined as solutions of the time independent equation:
\begin{equation}
\label{eq:spatial} \hat
R(x,E,i\epsilon\partial_{x})\psi_{\varepsilon}(x,E)=0.
\end{equation}
For any $E\in\Delta$, the set of such solutions is
$md$-dimensional.\smallpagebreak

We define:
\begin{equation}
\label{eq:nj} \forall l\in\{1,\dots, m\}\quad
N_{l}(x,E)=\sum\limits_{n=0}^{r}A_{ln}(x)E^{n}
\end{equation}
so that
$$
R(x, E, k)=\sum_{l=0}^m N_{l}(x,E)k^l .
$$
We first prove the following result:
\begin{Le}
We assume that $R$ and $\Delta$ satisfy  $\mathbf{(H1)}$ and
$\mathbf{(H3)}$. We have the following properties.
\begin{enumerate}
\item For any $E\in\Delta$ and any $x\in\R$, $N_{m}(x,E)$ is
invertible.
\item For $j\in\{1,\dots, m\}$, $(z,E)\mapsto N_{j}(z,E)$ is analytic
in $\rho_{Y}\times \Delta$.
\item If we define $H(x,E)$ by:
\begin{equation}
\label{eq:hamil}
H(x,E)=\left[\begin{array}{ccccc}0&Id&0&\dots&0\\0&0&Id&\dots&0\\
\vdots&\vdots&\vdots&\vdots&\vdots \\0&0&0&\dots&Id
\\-(N_{m}^{-1}N_{0})(x,E)&-(N_{m}^{-1}N_{1})(x,E)&-(N_{m}^{-1}N_{2})(x,E)&\dots&-(N_{m}^{-1}N_{m-1})(x,E)
\end{array}\right],
\end{equation}
then $\sigma(H(x,E))=\{k;\
\det(\sum\limits_{l=0}^{m}N_{l}(x,E)k^{l})=0\}=\{k;\ \det(R(x, E,
k))=0\}.$
\item The functions $\{(x,E)\mapsto k_{j}(x,E)\}_{j\in\{1,\dots,
md\}}$ are analytic in $\R\times\Delta$.
\item Ker $(R(x, E, k_j(x, E))))$ is one-dimensional.
\end{enumerate}
\end{Le}
\demo The singular values of $R(x, E,
k)=\sum_{l=0}^{m}N_{l}(x,E)k^{l}$ are the roots of the polynomial
\begin{equation}
\label{eq:charpol}
L(k)=\det\left(\sum_{l=0}^{m}N_{l}(x,E)k^{l}\right).
\end{equation}
 This
polynomial is of degree $md$ and the highest coefficient is $\det
N_{m}(x,E)$. According to $\mathbf{(H3)}$, since $L$ has $md$
distinct roots, $\det N_{m}(x,E)\neq 0$, which proves
(1).\smallpagebreak Assertion (2) is immediate, consider (3). A
complex number $k\in\sigma(H(x,E))$ if there exists
$\Phi\in\C^{md}\backslash\{0\}$ such that $H\Phi=k\Phi$. Block by
block computations show that $\Phi$ is of the form:
\begin{equation}
\label{eq:vect}
\Phi=\left(\begin{array}{c}\varphi\\k\varphi\\\vdots\\k^{m-1}
\varphi
\end{array}\right), \ \ \ \varphi \in \C^d,
\end{equation}
with $\det(\sum_{l=0}^{m}N_{l}(x,E)k^{l})= 0$ and
$\varphi\in\textrm{Ker}(\sum_{l=0}^{m}N_{l}(x,E)k^{l})$.\smallpagebreak
Again by \cite{Ka},  $\mathbf{(H3)}$ with $\delta>0$ implies that
the functions $k_{j}(x,E)$ are analytic in a complex neighborhood
of $\mathbb R \times \Delta$
 which proves (4). Point
(5) follows from (\ref{eq:vect}) and the fact that
$\sigma(H(x,E))$ is simple \smallpagebreak We introduce some
normalized eigenvectors of $R(x,E,k_{j}(x,E))$.
\subsection{Canonical eigenvectors of $R(x,E,k_{j}(x,E))$}
For a matrix $A$, we denote its adjoint by
$A^{*}=^{t}\overline{A}$.\smallpagebreak Fix $j\in\{1,\dots,
md\}$. Under $\mathbf{(H3)}$ and according to \cite{Ka}, we know
that there exist two vector valued functions $\xi_{j}$ and
$\xi_{j}^{\dag}$ with values in $\C^{d}$ such that:
\begin{enumerate}
\item The functions $(x,E)\mapsto\xi_{j}(x,E)$ and
$(x,E)\mapsto\xi_{j}^{\dag}(x,E)$ are analytic on
$\R\times\Delta$.
\item $\forall(x,E)\in\R\times\Delta,\quad\xi_{j}(x,E)\in\textrm{Ker}R(x,E,k_{j}(x,E)),\quad\xi_{j}^{\dag}(x,E)\in\textrm{Ker}R^{*}(x,E,k_{j}(x,E)).$
\end{enumerate}
\begin{Def}
The vector $\varphi_{j}=\alpha_{j}\xi_{j}$, with
\begin{equation}\label{eq:norm}
\alpha_{j}(x,E)=e^{-\int_{0}^{x}\frac{<\xi_{j}^{\dag},\partial_{k}R(u,E,k_{j})\partial_{x}\xi_{j}>+<\xi_{j}^{\dag},\frac{\partial_{x}k_{j}}{2}\partial^{2}_{k}R(u,E,k_{j})\partial_{x}\xi_{j}>}{<\xi_{j}^{\dag},\partial_{k}R(u,E,k_{j})\xi_{j}>}du}
\end{equation}
is called a canonical eigenvector associated to
$R(x,E,k_{j}(x,E))$.
\end{Def}
We notice the following facts:
\begin{itemize}
\item The vector $\varphi_{j}$ does not depend on the choice of
$\xi_{j}^{\dag}\in \mbox{Ker }(R^*(x,E,k_j(x,E))))$. In
particular, we can choose $\xi_{j}^{\dag}$ so that
$$<\xi_{j}^{\dag},\xi_{j}>=1.$$
\item  Condition \eqref{eq:norm} may seem
artificial but we shall see in the proof of Lemma \ref{le:decomp}
that it corresponds to the Kato normalization of the eigenvectors
of $H(x,E)$.
\end{itemize}
\subsection{Decomposition Lemma}

\begin{Le}\label{le:decomp}
We assume that $R$ and $\Delta$ satisfy $\mathbf{(H3)}$. Let
$\psi_{\varepsilon}(x,E)$ be a solution of~\eqref{eq:spatial}.
There exist $md$ functions $\{(z,E, \varepsilon)\mapsto c_{j}(z,E,
\varepsilon)\}_{j\in\{1,\dots, md\}}$ such that:
\begin{enumerate}
\item The function $\psi_{\varepsilon}(x,E)$ satisfies:
$$\forall l\in\{0,\dots, m-1\},\quad
(i\varepsilon\partial_{x})^{l}\psi_{\varepsilon}(x,E)=\sum\limits_{j=1}^{md}c_{j}(x,E,\varepsilon)k_{j}^{l}(x,E)e^{-\frac{i}{\varepsilon}\int_{0}^{x}k_{j}(y,E)dy}\varphi_{j}(x,E).$$
\item If we define
$$c(x,E,\varepsilon)=\left(\begin{array}{c}
c_{1}(x,E,\varepsilon)\\ \vdots\\c_{md}(x,E,\varepsilon)
\end{array}\right),$$
The vector $c$ satisfies the following differential equation:
\begin{equation}
\label{eq:diff_c}
\partial_{x}c(x,E,\varepsilon)=M(x,E,\varepsilon)c(x,E,\varepsilon),
\end{equation}
where the matrix $M$ is given by:
\begin{equation}\label{eq:mat_m}
M_{jl}(x,E)=a_{jl}(x,E)e^{i\frac{\Delta_{jl}(x,E)}{\varepsilon}},
\end{equation}
$$\mbox{with} \ \ \ \Delta_{jl}(x,E)=\int_{0}^{x}[k_{j}(u,E)-k_{l}(u,E)]du,$$
$$\mbox{and} \ \ \ \forall j\in\{1,\dots, md\},\quad a_{jj}(x,E)=0,$$
\begin{eqnarray}\label{eq:mat_a}
\forall j\neq l,\quad
& &a_{jl}(x,E)=\frac{1}{k_{j}(x,E)-k_{l}(x,E)}\times \\
&
&\nonumber\left[\frac{<\varphi_{j}^{\dag},R(x,E,k_{l})\partial_{x}
\varphi_{l}>+\partial_{x}k_{l}<\varphi_{j}^{\dag},[\partial_{k}R(x,E,k_{l})-
\partial_{k}R(x,E,k_{j})]\varphi_{l}>}{<\varphi_{j}^{\dag},
\partial_{k}R(x,E,k_{j})\varphi_{j}>}\right].
\end{eqnarray}
\item For any $j\in\{1,\dots, md\}$, $\varphi_{j}$ is a canonical
eigenvector of $R(x,E,k_{j}(x,E))$ and $\varphi_{j}^{\dag}$ is any
eigenvector in $\textrm{Ker}R^{*}(x,E,k_{j}(x,E))$.
\end{enumerate}
\end{Le}
{\bf Remark:} The set $\{\varphi_j\}_{j\in\{1,\cdots,md\}}$ is a
linearly dependent family of vectors in $\C^d$. The decomposition
in point (1) above corresponds to the familiar BKW Ansatz in
semiclassical analysis, see e.g. \cite{Fe:87}. \demo Let
$\psi_{\varepsilon}(x,E)$ be a solution of~\eqref{eq:spatial}. We
define:
\begin{equation}\label{eq:dec}\Psi_{\varepsilon}(x,E)=\left[\begin{array}{c}\psi_{\varepsilon}(x,E)\\(i\varepsilon\partial_{x})\psi_{\varepsilon}(x,E)\\\vdots\\(i\varepsilon\partial_{x})^{m-1}\psi_{\varepsilon}(x,E)\end{array}\right].
\end{equation}
Then $\Psi_{\varepsilon}(x,E)$ satisfies:
\begin{equation}\label{eq:big_vect}
i\varepsilon\partial_{x}\Psi_{\varepsilon}(x,E)=H(x,E)\Psi_{\varepsilon}(x,E).
\end{equation}
Equation~\eqref{eq:big_vect} has been studied in
\cite{joye-pfister95}, \cite{joye97}. We use the results obtained
there and write:
\begin{equation}\label{eq:bis}
H(x,E)=\sum\limits_{j=1}^{md}k_{j}(x,E)P_{j}(x,E),
\end{equation}
where the matrix valued functions $P_{j}(x,E)$ are the
one-dimensional eigenprojectors of $H(x,E)$ and satisfy:
$$\sum\limits_{j=1}^{md}P_{j}(x,E)=I_{md}.$$
Hypothesis ($\mathbf {H3}$) implies the existence of a basis of
eigenvectors of $H(x,E)$: $\{\Phi_{j}(x,E)\}_{j=1,\dots, md}$.
\smallpagebreak We determine these eigenvectors uniquely (up to a
constant depending on $E$) by requiring them to satisfy:
\begin{eqnarray}
H(x,E)\Phi_{j}(x,E)=k_{j}(x,E)\Phi_{j}(x,E),\quad\forall
j=1,\dots, md,\\
\label{eq:norm2}
P_{j}(x,E)\partial_{x}\Phi_{j}(x,E)=0,\quad\forall j=1,\dots, md.
\end{eqnarray}
Indeed, recall that if $W(x,E)$ is the solution of
\begin{equation}\label{wan}
\partial_{x}W(x,E)=\sum\limits_{j=1}^{md}(\partial_{x}P_{j}(x,E))
P_{j}(x,E)W(x,E),\quad W(0,E)=I_{md},
\end{equation} it is well known that $W(x,E)$ satisfies the
intertwining identity
$$W(x,E)P_{j}(0,E)=P_{j}(x,E)W(x,E),\quad\forall j\in\{1,\dots, md\}.$$
The generator of (\ref{wan}) being analytic in $E$, $W$ is
analytic in both variables $(x,E)\in \mathbb R\times \Delta$, see
\cite{dieudonne} section XI.5. Hence,
$$\Phi_j(x,E):= W(x,E)\Phi_j(0,E), \quad\forall j\in\{1,\dots, md\}$$
where $\{\Phi_j(0,E)\}_{j\in\{1,\cdots, md\}}$ is basis of
analytic eigenvectors of $H(0,E)$, satisfy
$$
P_j(x,E)\Phi_j(x,E)=\Phi_j(x,E) \quad \mbox{and eq.
(\ref{eq:norm2}).}
$$
We refer to \cite{Ka,joye-pfister95, joye97} for the
details.\smallpagebreak We will rewrite the eigenprojectors as:
$$P_{j}(x,E)=\frac{1}{<\Phi_{j}^{\dag}(x,E),\Phi_{j}(x,E)>}|\Phi_{j}(x,E)>
<\Phi_{j}^{\dag}(x,E)|,$$ where $\Phi_{j}^{\dag}(x,E)
\in\textrm{Ker}(H^{*}(x,E)-k_j(x,E))$, since $k_j(x,E)=
\overline{k_j}(x,E)$. \smallpagebreak We use the same notation for
duality in $\C^{m}$ and $\C^{md}$ since no confusion should arise.

Let us begin by specifying equation \eqref{eq:norm2} in our case.
We consider an eigenvector $\Xi_{j}(x,E)$ of $H(x,E)$. $\Xi_{j}$
is written as:
$$\Xi_{j}(x,E)=\left[\begin{array}{c}\xi_{j}(x,E)\\k_{j}(x,E)\xi_{j}(x,E)\\\vdots\\
k_{j}^{m-1}(x,E)\xi_{j}(x,E)\end{array}\right],\textrm{ with }
\xi_{j}(x,E)\in\textrm{Ker}R(x,E,k_{j}(x,E)).$$ The vector
$\Phi_{j}$ must be of the form $\Phi_{j}=\alpha_{j}\Xi_{j}$, where
$\alpha_j\in\C$ and we define $\varphi_{j}=\alpha_{j}\xi_{j}$.
Then:
$$\Phi_{j}=\left[\begin{array}{c}\varphi_{j}\\k_{j}\varphi_{j} \\
\vdots\\k_{j}^{m-1}\varphi_{j}\end{array}\right].$$ Now, if
$\Xi^\dag_j(x,E)\in$ Ker $(H^*(x,E)-k_j(x,E))$ then $\Phi_{j}$
satisfies \eqref{eq:norm2} if:
$$\frac{\partial_{x}\alpha_{j}}{\alpha_{j}}=-\frac{<\Xi_{j}^{\dag},\partial_{x}\Xi_{j}>}{<\Xi_{j}^{\dag},\Xi_{j}>}.$$
It remains to choose $\Xi^\dag_j$ and to compute
$<\Xi_{j}^{\dag},\partial_{x}\Xi_{j}>$ and
$<\Xi_{j}^{\dag},\Xi_{j}>$.\smallpagebreak We start with the
computation of the vector $\Xi_{j}^{\dag}(x,E)$. It is an
eigenvector of $H^{*}(x,E)$ associated with the eigenvalue
$\overline{k_{j}}(x,E)=k_{j}(x,E)$. Let $\xi_j^{\dag}(x,E) \in $
Ker$(R^*(x,E,k_j(x,E)))$. We check that we can take
$$\Xi_{j}^{\dag}=\left[\begin{array}{c}\sum\limits_{l=1}^{m}k_{j}^{l-1}N_{l}^{*}\xi_{j}^{\dag}\\\sum\limits_{l=2}^{m}k_{j}^{l-2}N_{l}^{*}\xi_{j}^{\dag}\\\vdots\\N_{m}^{*}\xi_{j}^{\dag}\end{array}\right].$$
Then:
$$<\Xi_{j}^{\dag},\Xi_{j}>=\sum\limits_{p=1}^{m}k_{j}^{p-1}\sum\limits_{l=p}^{m}k_{j}^{l-p}<N_{l}^{*}\xi_{j}^{\dag},\xi_{j}>=\sum\limits_{l=1}^{m}k_{j}^{l-1}\sum\limits_{p=1}^{l}<N_{l}^{*}\xi_{j}^{\dag},\xi_{j}>=<\xi_{j}^{\dag},\sum\limits_{l=1}^{m}l k_{j}^{l-1}N_{l}\xi_{j}>$$
$$=<\xi_{j}^{\dag},\partial_{k}R(x,E,k_{j}(x,E))\xi_{j}>.$$
Similarly, we compute:
$$<\Xi_{j}^{\dag},\partial_{x}\Xi_{j}>=<\xi_{j}^{\dag},\partial_{k}R(x,E,k_{j})\partial_{x}\xi_{j}>+<\xi_{j}^{\dag},\frac{\partial_{x}k_{j}}{2}\partial^{2}_{k}R(x,E,k_{j})\partial_{x}\xi_{j}>.$$

This implies that $\varphi_{j}$ is a canonical eigenvector of
$R(x,E,k_{j})$. \smallpagebreak From \cite{joye-pfister95,
joye97}, we know that any solution to~\eqref{eq:big_vect} can be
written as:
$$\Psi_{\varepsilon}(x,E)=\sum\limits_{j=1}^{md}c_{j}(x,E,\varepsilon)e^{-\frac{i}{\varepsilon}\int_{0}^{x}k_{j}(y,E)dy}\Phi_{j}(x,E),$$
where the scalar coefficients $c_{j}$ satisfy the differential
equation: $\partial_{x}c=Mc,$ where $M$ is given
by~\eqref{eq:mat_m}, and
$$
a_{jl}=-\frac{<\Phi_{j}^{\dag},\partial_{x}\Phi_{l}>}{<\Phi_{j}^{\dag},\Phi_{j}>}.$$
We compute
\begin{eqnarray}<\Phi_{j}^{\dag},\partial_{x}\Phi_{l}>&=&\partial_{x}k_{l}\sum_{p=2}^{m}
\sum_{q=p}^{m}(q-1)k_{j}^{q-p}k_{l}^{p-2}<N_{q}^{*}\varphi_{j}^{\dag},\varphi_{l}>\nonumber
\\
&+&\sum_{p=1}^{m}\sum_{q=p}^{m}k_{j}^{q-p}k_{l}^{p-1}<N_{q}^{*}\varphi_{j}^{\dag},\partial_{x}\varphi_{l}>.
\end{eqnarray} By interchanging the indices $p$ and $q$ and
according to the formula
$$\forall a\neq b,\qquad\sum_{p+l=s}a^{p}b^{l}=\frac{a^{s+1}-b^{s+1}}{a-b},$$
we obtain formula \eqref{eq:mat_a}. The first statement of the
Lemma stems from formula (\ref{eq:dec}). This ends the proof of
Lemma~\ref{le:decomp}.
\subsection{Behavior of the matrix $M$}
The following Lemma describes the behavior of the coefficients
$a_{ij}$ and phases entering the definition of $M$.
\begin{Le}
\label{le:beh_a}We assume that ({\bf H1}), ({\bf H2}) and ({\bf
H3}) are satisfied. Then,
\begin{itemize}
\item The eigenvalues $k_{j}$ satisfy for any $k\in\N$ and any
$l\in\N$:
\begin{equation}\label{eq:deck}
\forall
E\in\Delta,\quad\sup\limits_{x\rightarrow+\infty}|x|^{2+\nu}|\partial_{E}^{l}\partial_{x}^{k}(k_{j}(x,E)-k_{j}(+\infty,E))|+\sup\limits_{x\rightarrow-\infty}|x|^{2+\nu}|\partial_{E}^{l}\partial_{x}^{k}(k_{j}(x,E)-k_{j}(-\infty,E))|<\infty.
\end{equation}
\item The eigenvectors $\varphi_{j}$ satisfy for any
$l\in\N$, uniformly in $E\in\Delta$:
\begin{equation} \label{eq:decphi1}
\sup\limits_{x\rightarrow+\infty}|x|^{1+\nu}\|\partial_{E}^{l}(\varphi_{j}(x,E)-\varphi_{j}(+\infty,E))\|+\sup\limits_{x\rightarrow-\infty}|x|^{1+\nu}\|\partial_{E}^{l}(\varphi_{j}(x,E)-\varphi_{j}(-\infty,E))\|<\infty.
\end{equation}
\item Moreover, for any $k\in\N^*$ and $l\in\N$, uniformly in $E\in\Delta$:
\begin{equation} \label{eq:decphi}
\sup\limits_{x\rightarrow+\infty}|x|^{2+\nu}\|\partial_{E}^{l}\partial_{x}^{k}(\varphi_{j}(x,E)-\varphi_{j}(+\infty,E))\|+\sup\limits_{x\rightarrow-\infty}|x|^{2+\nu}\|\partial_{E}^{l}\partial_{x}^{k}(\varphi_{j}(x,E)-\varphi_{j}(-\infty,E))\|<\infty.
\end{equation}
\item For  any $k\in\N$ and any $l\in\N$, the coefficients of the matrix $M$ satisfy
uniformly in $E\in\Delta$:
\begin{equation} \label{eq:deca} \forall x\in\R, \quad\forall
(j,p)\in\{1,\dots, md\}^{2},\quad
|\partial_{E}^{l}\partial_{x}^{k}a_{jp}(x,E)||x|^{2+\nu}<\infty.\end{equation}
\item Let
\begin{equation} \label{d1}\omega_{j}(\pm\infty,E)=\int_{0}^{\pm\infty}[k_{j}(y,E)-
k_{j}(\pm\infty,E)]dy,\end{equation} and
\begin{equation} \label{d2}\int_{0}^{x}k_{j}(y,E)dy=x
k_{j}(\pm\infty,E)+\omega_{j}(\pm\infty,E)+r_{j}^{\pm}(x,E).\end{equation}
Then we have, uniformly in  $E\in\Delta$, and for any $n\in\N$,
\begin{equation} \label{d3}
\forall j\in\{1,\dots,md\},\quad \sup\limits_{x>0}|x|^{1+\nu}|
\partial_{E}^{n}r_{j}^{+}(x,E)|+\sup\limits_{x<0}|x|^{1+\nu}|
\partial_{E}^{n}r_{j}^{-}(x,E)|<\infty.\end{equation}
\end{itemize}
\end{Le}
We prove Lemma \ref{le:beh_a} in section \ref{sec:techn3}.
\subsection{The vector $c$}
\label{sec:vector_c} In the following lemma, we describe the
behavior of the vector $c$ defined by the ODE (\ref{eq:diff_c}).
\begin{Le}
We assume that $\mathbf{(H1)}$, $\mathbf{(H2)}$, and
$\mathbf{(H3)}$ are satisfied.
\begin{itemize}
\label{le:as}
\item For any $E\in\Delta$ and $\epsilon>0$, the limits $c_{j}(\pm\infty,E,\varepsilon)$
exist for all $j=1,\cdots, md$.
\item If the initial conditions to (\ref{eq:diff_c})
are chosen so that $c(-\infty,E,\varepsilon)$ is uniformly bounded
in $E\in\Delta$ and $\epsilon>0$, then we have for some constant
$C$ uniform in $\varepsilon$ and $E\in\Delta$:
$$|\partial_{E}c_{j}(\pm\infty,E,\varepsilon)|+|c_{j}(\pm\infty,E,\varepsilon)|<C,$$
$$ \sup\limits_{x>0}|x|^{\nu}|\partial_{E}c_{j}(x,E,\varepsilon)-\partial_{E}c_{j}
(+\infty,E,\varepsilon)|+\sup\limits_{x<0}|x|^{\nu}|\partial_{E}c_{j}(x,E,\varepsilon)-
\partial_{E}c_{j}(-\infty,E,\varepsilon)|<C,$$
$$\sup\limits_{x>0}|x|^{1+\nu}|c_{j}(x,E,\varepsilon)-c_{j}(+\infty,E,\varepsilon)|
+\sup\limits_{x<0}|x|^{1+\nu}|c_{j}(x,E,\varepsilon)-c_{j}(-\infty,E,\varepsilon)|<C.$$
\end{itemize}
\end{Le}
\noindent{\bf Remarks:}\\
i) As the proof shows, the condition
$\sup\limits_{E\in\Delta\atop\varepsilon\rightarrow 0
}\|c(-\infty,E,\varepsilon)\|<\infty$ can  be replaced by
\begin{equation}\label{ccb}
\exists x_0\in\R \ \ \mbox{such that} \ \
\sup_{E\in\Delta\atop\varepsilon\rightarrow 0 }
\|c(x_0,E,\varepsilon)\|<\infty.
\end{equation}
ii) In the construction of solutions to (\ref{evol}) by means of
an energy density, we can (and will) always assume that the
initial conditions, wherever they are chosen, are uniformly
bounded in energy:
\begin{equation}\label{ccbe}
\exists x_0\in\R \ \ \mbox{such that} \ \ \sup_{E\in\Delta }
\|c(x_0,E,\varepsilon)\|<\infty.
\end{equation}
iii) The equation being linear, we can actually always assume
condition
(\ref{ccb}) holds. This is what we do in the rest of the paper.\\

We shall prove Lemma~\ref{le:as} in section
\ref{sec:techn2}.\smallpagebreak According to Lemma~\ref{le:as},
we can define the stationary scattering matrix $S(E,\varepsilon)$
by:
\begin{equation}\label{eq:scat}
S(E,\varepsilon)c(-\infty,E,\varepsilon)=c(+\infty,E,\varepsilon).
\end{equation}
In order to describe the time-dependent scattering processes we
are interested in, we need more detailed informations about the
stationary $S$-matrix.

\section{Complex BKW analysis}

\label{sec:wkb} In this section, the parameter $\delta>0$ is still
kept fixed. All the information about transmissions and
transitions among the asymptotic eigenstates is contained in the
asymptotic values of the coefficients $c_{j}(x,E,\pm\infty)$
defined in section \ref{sec:vector_c} and hence in the stationary
scattering matrix $S(E,\varepsilon)$. We extract this information
by mimicking the complex BKW method of \cite{joye-pfister95} and
\cite{joye97}, while keeping track of the $E$-dependence.
\smallpagebreak In the simplest setting, the complex BKW method
requires hypotheses on the behavior of the so-called Stokes-lines
for equation \eqref{eq:big_vect} in order to provide the required
asymptotics. These hypotheses are global in nature, and in general
are extremely difficult to check. See {\it e.g.}, \cite{Fe:87,
Fe:89}. However, in the physically relevant situation of avoided
crossings, they can be easily checked, as is proven in
\cite{joye97} and will be recalled in the next Section. We
restrict our attention to these avoided crossing
situations.\smallpagebreak To study the $S$-matrix, it is enough
to consider the coefficients $c_{j}$ that are uniquely defined by
the conditions
\begin{equation} \label{inco} c_{j}(-\infty, E,\varepsilon)=1
\quad c_{k}(-\infty, E,\varepsilon)=0,\ \ \ \mbox{for all}\ \ \
k\neq j.
\end{equation}
The key of the complex BKW method lies in the multivaluedness of
the eigenvalues and the eigenvectors of the analytic generator
$H(x,E)$ in the complex $x$ plane.\smallpagebreak According to
$\mathbf{(H3)}$, the eigenvalues and eigenvectors of $H(x,E)$ are
analytic in $x$ on the real axis. They may have branch points in
$\rho_{Y}$ that are located in
\begin{equation}
\label{eq:omega} \Omega(E)=\{z\in\rho_Y\ ;\ \exists\ j\neq l\quad
\mbox{such that}\quad k_{j}(z,E)=k_{l}(z,E)\}\end{equation}
\subsection{The set $\Omega(E)$}
By the Schwarz reflection principle, for any $E\in\Delta$, we have
$\overline{\Omega}(E)=\Omega(E)$. Besides, the set
$\bigcup\limits_{E\in\Delta}\Omega(E)$ is bounded in $\rho_Y$.

\smallpagebreak We have the following description of $\Omega(E)$,
see \cite{joye97}:
\begin{Le}
Fix $E_{0}\in\Delta$. There exists a neighborhood $\Delta_{0}$ of
$E_{0}$ and a finite number $R$ of bounded open sets
$\{\Omega_{i}\}_{i\in\{1, \dots, R\}}$ in $\rho_Y\cap\C_{+}$ such
that:
\begin{itemize}
\item For any $E\in\Delta_{0}$,
$\Omega(E)\subset\bigcup_{1}^{R}\Omega_{i}\bigcup_{1}^{R}\overline{\Omega_{i}}$.
\item For all $i\in\{1,\dots, R\}$, $\Omega_{i}\cap\R=\emptyset$.
\item For any $E\in\Delta_{0}$, and $i\in\{1,\dots, R\}$, $\Omega_{i}$ contains only one
crossing point. This point is a crossing point for finitely many
distinct couples of modes.
\end{itemize}
\end{Le}
We define
$\Omega=\bigcup_{1}^{R}\Omega_{i}\bigcup_{1}^{R}\overline{\Omega_{i}}$.
\smallpagebreak Under our genericity hypotheses, we have the
following local behavior at a complex crossing point
$z_{0}\in\Omega(E_{0})$,:
$$
k_{j}(z,E_{0})-k_{l}(z,E_{0})\simeq\gamma(E_{0})(z-z_{0})^{1/2}(1+O(z-z_{0})).$$
The eigenprojectors of $H(x,E)$ also admit multivalued extensions
in $\rho_Y\setminus\Omega(E)$, but they diverge at generic
eigenvalue crossing points. We only have to deal with generic
crossing points. To see what happens to a multivalued function $f$
in $\rho_Y\setminus\Omega$ when we turn around a crossing point,
we adopt the following convention: For $E$ fixed, we denote by
$f(z,E)$ the analytic continuation of $f$ defined in a
neighborhood of the origin along some path from $0$ to $z$. Then
we perform the analytic continuation of $f(z, E)$ along a
negatively oriented loop that surrounds only one connected
component $\Omega_{i}$ of $\Omega$. We denote by $\tilde{f}(z,E)$
the function we get by coming back to the original point $z$. We
define $\zeta_0$ to be a negatively oriented loop, based at the
origin, that encircles only $\Omega_{i}$ when
$\Omega_{i}\in\C_{+}$. When $\Omega_{i}\in\C_{-}$, we choose
$\zeta_0$ to be positively oriented.\smallpagebreak We now fix
$\Omega_{i}\in\C_{+}$. For any $E\in\Delta_{0}$, if we
analytically continue the set of eigenvalues
$\{k_j(z,E)\}_{j=1}^{md}$, along a negatively oriented loop around
$\Omega_{i}$, we get the set $\{\widetilde{k}_j(z,E)\}_{j=1}^{md}$
with $$
  \widetilde{k}_j(z,E)=k_{\pi_0(j)}(z,E),\quad\mbox{for}\quad j=1,\cdots,md,
$$ where \begin{equation}\label{eq:pi0}
  \pi_0 :\;\{1,2,\cdots,m\}\rightarrow\{1,2,\cdots,md\}
\end{equation} is a permutation that depends on $\Omega_{i}$.
 As a consequence, the
eigenvectors $\Phi_{j}$ possess multi-valued analytic extensions
in $\rho_Y\backslash \Omega$. The analytic continuation
$\widetilde{\Phi}_j(z,E)$ of $\Phi_j(z,E)$ along a negatively
oriented loop around $\Omega_{i}$, must be proportional to
${\Phi}_{\pi_0(j)}(z,E)$. Thus, for $j=1,2,\cdots,md$, there
exists $\theta_j(\zeta_0)\in\C$, such that
\begin{equation}\label{eq:theta} \widetilde{\Phi}_j(z,E)\ =\
e^{-i\theta_j(\zeta_0,E)}{\Phi}_{\pi_0(j)}(z,E). \end{equation}

\smallpagebreak

The above implies a key identity for the analytic extensions of
the coefficients $c_j(z,E,\varepsilon)$, $z\in \rho_Y\backslash
\Omega$. Since the solutions to (\ref{eq:big_vect}) are analytic
for all $z\in\rho_Y$, the coefficients $c_j$ must also be
multi-valued. In our setting, Lemma 3.1 of \cite{joye97} implies
the following lemma.
\begin{Le}\label{le:echco}
For any $j=1,\dots,md$, we have \begin{equation}\label{eq:echco}
\widetilde{c}_j(z,E,\varepsilon)\
e^{-\,i\,\int_{\zeta_0}\,k_j(u,E)\,du/\varepsilon}\
e^{-\,i\,\theta_j(\zeta_0,E)}\ =\ c_{\pi_0(j)}(z,E,\varepsilon)
\end{equation}
where $\zeta_0$ and $\pi_0(j)$ are defined as above and are
independent of $E\in\Delta_{0}$.
\end{Le}
{\bf Remark:} Since $\Omega$ has a finite number of connected
components, it is straightforward to generalize the study of the
analytic continuations around one crossing point to analytic
continuations around several crossing points. The loop $\zeta_0$
can be rewritten as a concatenation of finitely many individual
loops, each encircling only one connected component of $\Omega$.
The permutation $\pi_0$ is given by the composition of associated
permutations. The factors $e^{-i\theta_j(\zeta_0,E)}$ in
(\ref{eq:theta}) are given by the product of the factors
associated with the individual loops. The same is true for the
factors
$\exp\left(\,-\,i\,\int_{\zeta_0}\,k_j(z,E)\,dz/\varepsilon\,\right)$
in Lemma \ref{le:echco}. \smallpagebreak

\subsection{Dissipative domains} We now describe how to use the above properties
in order to control the limit $\varepsilon\rightarrow 0$. The
details may be found in \cite{joye97}.

The idea is to integrate the integral equation corresponding to
(\ref{eq:diff_c}) along paths that go above (or below) one or
several crossing points, and then to compare the result with the
integration performed along the real axis. As $z\rightarrow-\infty
$ in $\rho_Y$ these paths become parallel to the real axis so that
the coefficients take the same asymptotic value
${c}_m(-\infty,E,\varepsilon)$ along the real axis and the
integration paths. Since the solutions to (\ref{eq:big_vect}) are
analytic, the results of these integrations must agree as $\Re
z\rightarrow\infty$. Therefore, (\ref{eq:echco}) taken at
$z=\infty$ yields the asymptotics of $c_{\pi_0(j)}(\infty,E,
\varepsilon)$, provided we can control
$\widetilde{c}_j(z,E,\varepsilon)$ in the complex plane. We argue
below that this can be done in the so-called dissipative domains
of the complex plane. We do not go into the details of these
notions because a result of \cite{joye97} will enable us to get
sufficient control on $\widetilde{c}_j(z,E,\varepsilon)$ in the
avoided crossing situation, to which we restrict our attention.

We recall that $\Delta_{jl}$ is defined in \eqref{eq:mat_m}. We
rewrite \eqref{eq:diff_c} as an integral equation:
\begin{equation}
\label{eq:diff_d}
c_{j}(x,E,\varepsilon)=c_{j}(x_{0},E,\varepsilon)+\int_{x_{0}}^{x}\sum\limits_{l}a_{jl}(x',E)e^{\frac{i\Delta_{jl}(x',E)}{\varepsilon}}c_{l}(x',E,\varepsilon)dx'.
\end{equation}
By explicit computation, we check that (\ref{eq:diff_d}) can be
extended to $\rho_Y\setminus\Omega$. We integrate by parts in
(\ref{eq:diff_d}) to see that (\ref{eq:diff_d}) with
$x_{0}=-\infty$ can be rewritten as:
\begin{eqnarray}\label{eq:intpp}\hskip -12pt
  \widetilde{c}_m(z,E,\varepsilon)&=&\delta_{jm}
  -i\varepsilon\sum_{l}\,
  \frac{\widetilde{a}_{ml}(z,E)}{\widetilde{k}_m(z,E)-
  \widetilde{k}_l(z,E)}\,
  e^{i\widetilde{\Delta}_{ml}(z,E)/\varepsilon}\,
  \widetilde{c}_l(z,E,\varepsilon)
\\
  &&+\ i\varepsilon^2\sum_{l}\int_{-\infty}^z
  {\left(\frac{\partial}{\partial z'}\,
  \frac{\widetilde{a}_{ml}(z',E)}
{\widetilde{k}_m(z',E)-\widetilde{k}_l(z',E)}\right)}\,
  e^{i\widetilde{\Delta}_{ml}(z',E)/\varepsilon}\,
  \widetilde{c}_l(z',E,\varepsilon)\,dz'\nonumber
\\
  &&+i\varepsilon\sum_{l,p}\,\int_{-\infty}^z
  \frac{\widetilde{a}_{ml}(z',E)\,
      \widetilde{a}_{lp}(z',E)}
  {\widetilde{k}_m(z',E)-\widetilde{k}_l(z',E)}\,
  e^{i\widetilde{\Delta}_{mp}(z',E)/\varepsilon}\,
  \widetilde{c}_p(z',E,\varepsilon)\,dz',\nonumber
\end{eqnarray}
as long as the chosen path of integration does not meet $\Omega$.
Here, $\widetilde{\phantom{c}}$ denotes the analytic continuation
along the chosen path of integration of the corresponding function
defined originally on the real axis. This distinguishes
$\widetilde{c}_m(\infty,E,\varepsilon)$ from
${c}_m(\infty,E,\varepsilon)$ computed along the real axis as
$x\rightarrow\infty$. These quantities may differ since the
integration path may pass above (or below) points of $\Omega$. If
the exponential factors in (\ref{eq:intpp}) are all uniformly
bounded when $\varepsilon\rightarrow 0$, as it is the case when
the integration path coincides with the real axis, it is
straightforward to get bounds of the type
\begin{equation}\label{eq:ifdis} {c}_m(z,E,\varepsilon)\ =\
\delta_{jm}\ +\ O_E(\varepsilon).
\end{equation}
However, when dealing with $\tilde{c}_m$ in the complex plane,
these exponential factors are usually not uniformly bounded and
one needs to restrict integration paths to certain domains in
which useful estimates
can be obtained: \\

One defines a {\em Dissipative domain} for index $j$, $D_j\subset
\rho_Y\setminus \Omega$ associated with the initial condition
(\ref{inco}), by the conditions:
\begin{itemize}
\item $D_j\subset \rho_Y\setminus \Omega$ and $\sup_{z\in D_j}\re z=\infty,
\inf_{z\in D_j}\re z=-\infty$,
\item For any $z\in D_j$ and any index $k\in\{1,\dots, md\}$, there exists a
path $\gamma^k\subset D_j$, parametrized by $u\in (-\infty,t]$
which the regularity properties
$$\lim_{u\rightarrow -\infty }\re \gamma^k(u)=-\infty\ , \
\gamma^k(t)=z \  \mbox{ and }\ \sup_{z\in D_j}\sup_{u\in
(-\infty,t]}|\partial_u \gamma^k(u)|<\infty$$
\item  $\gamma^k$ satisfies the monotonicity properties
$$u\mapsto \im \widetilde{\Delta}_{jk}( \gamma^k(u))\ \mbox{ is non-decreasing
on }\ (-\infty,t].$$
\end{itemize}
Again, as it is well known, the existence of paths from $-\infty$
to $+\infty$ passing above (or below) points in $\Omega$ and along
which the exponentials can be controlled is difficult to check in
general. We can overcome these complications by restricting
attention to avoided crossing situations where the existence of
dissipative domains for all indices has been proven in
\cite{joye97}, see hypothesis {\bf (AC)} below. The interest of
the
definition above lies in the following property.\\

When a dissipative domains exists for the index $j$,
(\ref{eq:echco}) and (\ref{eq:ifdis}) imply
\begin{equation}\label{eq:wkb} c_{\pi_0(j)}(\infty,E,\varepsilon)\
=\ e^{-\,i\,\int_{\zeta_0}\,k_j(u,E)\,du/\varepsilon}\
e^{-\,i\,\theta_j(\zeta_0,E)}\ (1+O_E(\varepsilon)),
\end{equation} where the $O_E(\varepsilon)$ estimate is uniform
for $E\in\Delta_{0}$. This is the main result of Proposition 4.1
in \cite{joye97} for our purpose, under the assumption that a
dissipative domain $D_j$ exists.

In our context, all quantities depend on $E\in\Delta_{0}$.
However, by carefully following the proof of Proposition 4.1 of
\cite{joye97}, it is not difficult to check that the estimate
(\ref{eq:ifdis}) is uniform for $E\in\Delta_{0}$. For later
purposes, we also note here that under the same hypotheses on the
exponential factors, $\frac{\partial\phantom{E}}{\partial
E}\widetilde{c}_m(z,E,\varepsilon)$ is uniformly bounded for
$0<\varepsilon<\varepsilon_0$ and $E\in\Delta_{0}$ for some fixed
$\varepsilon_0$, by differentiation of (\ref{eq:intpp}). See the
proof of Lemma \ref{le:as} for this property on the real axis.

\subsection{Avoided crossings} We now
make use of the avoided crossing situation, that allows us to
prove the existence of dissipative domains. We thus restore the
parameter $\delta$ in the notation. We therefore work under {\bf
(H3)} and under the following assumption on the patterns of
crossings for the modes
 $\{k_{j}(x,E,0)\}$:\\

{\bf (AC)}:\begin{itemize}
\item For all $x<x_1(E)$,
$$
  k_1(x,E,0)< k_2(x,E,0)<\cdots <k_{md}(x,E,0).
$$
\item For all $j<l\in\{1,2,\cdots ,md\}$,
there exists at most one $x_r(E)$ with
$$
  k_j(x_r(E), E, 0)\,-\,k_l(x_r(E), E, 0)\ =\ 0,
$$
and if such an $x_r(E)$ exists, we have
\begin{equation}\label{gc}
\frac{\partial\phantom{i}}{\partial x}
  \left(k_j(x_r(E), E, 0)-k_l(x_r(E), E, 0)\right)\ >\ 0.
\end{equation}
\item For all $j\in\{1,2,\cdots ,md\}$,
the mode $k_j(x, E, 0)$ crosses modes whose indices are all
superior to $j$ or all inferior to $j$.
\end{itemize}

\smallpagebreak To any given pattern of real crossings for the
group $\{k_{j}(x,E,0)\}$, with $E\in\Delta_{0}$, we associate a
permutation $\pi$ as follows. The modes $\{k_{j}(x,E,0)\}$ are
labelled in ascending order at $x\simeq -\infty$, by
$\mathbf{(H3)}$. Since there are no real crossings for
$E\in\Delta$ and as $x\rightarrow +\infty$, the values
$\{k_{j}(x,E,0)\}$ are ordered uniformly in $E\in\Delta$ at
$x=+\infty$. If $k_{j}(+\infty,E,0)$ is the $k^{\textrm{th}}$
eigenvalue in ascending order at $x=+\infty$, the permutation
$\pi$ is defined by
\begin{equation}\label{eq:perm}
\pi(j)=k.
\end{equation}
Let $E$ be in a sufficiently small interval $\Delta_{0}$. For a
loop $\zeta_{0}$ that surrounds all the complex crossing points
and $\pi_{0}$ the associated permutation (see \eqref{eq:echco}),
$\pi_{0}$ corresponds to the permutation $\pi$. \smallpagebreak We
can now restate the main result of \cite{joye97} that describes
the asymptotics of the coefficients defined in \eqref{eq:intpp}.
We only have to check that, for small $\delta>0$, dissipative
domains exist and do not depend on $E\in\Delta_{0}$. We refer to
\cite{joye97} for the details. The construction of these
dissipative domains is based on a perturbation of the case
$\delta=0$.
By mimicking the arguments of \cite{joye97}, as in \cite{JH:04} we
obtain that estimates of the type \eqref{eq:wkb} are true for
certain indices $j$ and $n$, determined by the permutation
\eqref{eq:perm}:
\begin{Th}\label{th:PERCO}
Assume that $\mathbf{(H1)}$ to  $\mathbf{(H3)}$ are satisfied and
that $\mathbf{(AC)}$ holds. If $\delta>0$ and $\Delta_{0}$ are
small enough, the $\pi(j),j$ elements of the matrix
$S(E,\varepsilon)$, with $\pi(j)$ defined in \eqref{eq:perm} have
small $\varepsilon$ asymptotics for all $j=1,\cdots,md$ given by
$$
S_{\pi(j),j}(E,\varepsilon)\ =\
\prod_{l=j}^{\pi(j)\mp1}e^{-i\theta_l(\zeta_l,E,\delta)}\,
e^{i\int_{\zeta_l}\,k_l(z,E,\delta)\,dz/\varepsilon} \left(1+O_{E,
\delta}(\varepsilon)\right), \qquad\quad\pi(j)\ \left\{\,{ >j
\atop <j}\right.
$$
where, for $\pi(j)>j$ (resp. $\pi(j)<j$), $\zeta_l$,
$l=j,\cdots,\pi(j)-1$ (resp. $l=j,\cdots,\pi(j)+1$), denotes a
negatively (resp. positively) oriented loop based at the origin
which encircles the complex domain $\Omega_r$ (resp.
$\overline{\Omega_r}$) corresponding to the avoided crossing
between $k_l(x,E,\delta)$ and $k_{l+1}(x,E,\delta)$ (resp.
$k_{l-1}(x,E,\delta)$). The $\int_{\zeta_l}k_l(z,E,\delta)\,dz$
denotes the integral along $\zeta_l$ of the analytic continuation
of $k_l(0,E,\delta)$, and $\theta_l(\zeta_l,E, \delta)$ is the
corresponding factor defined by (\ref{eq:theta}).
\end{Th}

{\bf Remark:} Under our regularity hypotheses in $\delta$, it is
easy to get the following properties, see \cite{joye94},
$$\lim_{\delta\rightarrow 0}\int_{\zeta_l}\,k_l(z,E,\delta)\,dz =0.$$

Let us emphasize here that we do not have access to all
off-diagonal elements of the $S$-matrix; those we can
asymptotically compute are determined by the pattern of avoided
crossings. Moreover, there are cases in which one can compute all
elements of the $S$-matrix, due to supplementary symmetries in the
problem, see \cite{jopfi}. Sometimes, the coefficients to which we
have access are not even the largest ones in the avoided crossing
situation, as shown in \cite{jopfi}.

\smallpagebreak On the basis of steepest descent arguments,
transitions between modes that do not display avoided crossings,
{\it i.e.}, that are separated by a gap of order 1 as
$\delta\rightarrow 0$, are expected to be exponentially smaller
than the transitions we control by means of Theorem
\ref{th:PERCO}, as $\delta$ shrinks to zero. Since the
coefficients in the exponential decay rates given by the theorem
vanish in the limit $\delta\rightarrow  0$, it is enough to show
that the decay rates of the exponentially small transitions
between well separated levels are independent of $\delta$.

That is the meaning of the following proposition, which is proven
in \cite{JH:04}:

\begin{Pro}\label{pro:smaller} We assume that $\mathbf{(H3)}$ is satisfied.
Further assume that the eigenvalues of $H(x,E,\delta)$ can be
separated into two distinct groups $\sigma_1(x,E,\delta)$ and
$\sigma_2(x,E,\delta)$ that display no avoided crossing for
$E\in\Delta$, {\it i.e.}, such that
$$
\inf_{\delta\geq 0,E\in\Delta\atop x\in \rho_Y\cup\{\pm\infty\}}
\mbox{\em dist}(\sigma_1(x,E,\delta),\,\sigma_2(x,E,\delta))\
\geq\ g\
>\ 0.
$$
Let $P(x,E,\delta)$ and $Q(x,E,\delta)=\un -P(x,E,\delta)$ be the
projectors onto the spectral subspaces corresponding to
$\sigma_1(x,E,\delta)$ and $\sigma_2(x,E,\delta)$ respectively,
and let $U_\varepsilon(x,x_0,E,\delta)$ be the (space-) evolution
operator corresponding to the equation
\begin{equation}\label{eq:evol}
i\,\varepsilon\,\frac{d}{dx}\,U_\varepsilon(x,x_0,E,\delta)\ =\
H(x,E,\delta)\
U_\varepsilon(x,x_0,E,\delta),\qquad\mbox{with}\qquad
U_\varepsilon(x_0,x_0,E,\delta)=\un . \end{equation} Then, for any
$\delta>0$, there exists $\varepsilon_0(\delta)$, $C(\delta)>0$
depending on $\delta$, and $\Gamma >0$ independent of $\delta$,
such that for all $\varepsilon \leq \varepsilon_0(\delta)$,
$$
\lim_{x\rightarrow\infty \atop x_0\rightarrow -\infty}\
\|\,P(x,E,\delta)\,U_\varepsilon(x,x_0,E,\delta)\,
Q(x_0,E,\delta)\,\|\ \leq\ C(\delta)\ e^{-\Gamma/\varepsilon}.
$$
\end{Pro}
This Proposition implies that the stationary transitions between
modes without an avoided crossing are exponentially smaller than
transitions between modes displaying an avoided crossings. It also
shows that
in any case, these transitions are all exponentially small. \\

Let us end this section by remarking that we have always specified
initial conditions at $x=-\infty$. Obviously, the BKW analysis can
be equally performed for coefficients whose initial conditions are
specified at $x=+\infty$, {\it mutatis mutandis}.

\section{Exact solutions to the time-dependent equation}
\label{sec:t_depdt} In this Section, we construct solutions to
\begin{equation}\label{evolagain}
{\mathcal R}(x, i\varepsilon \partial_t, i\varepsilon \partial_x )
\phi(x,t,\varepsilon)=0, \quad x\in\R
\end{equation} by taking
time-dependent superpositions of the generalized eigenvectors
$\psi_{\varepsilon}(x,E)$, for $E\in\Delta$, studied in Section
\ref{sec:geneigen}. We investigate particularly these exact
solutions in the scattering regime of large but finite times $t$,
and for any fixed $\varepsilon>0$, not necessarily small.\\

The superpositions of generalized eigenvectors depend on an energy
density $Q(E,\varepsilon)$ that might be complex valued. We assume
that the following regularity conditions holds:
\begin{description}
\item[(C0)] The density $E\mapsto Q(E,\varepsilon)$ is supported on $\Delta$  and
is $C^{1}$ on $\Delta$, for any fixed $\varepsilon$.\\
Moreover, (\ref{ccbe}) is true.
\end{description}\vspace{.2cm}
In this Section, the parameter $\delta$ is fixed and we omit it in
the notations. We work under the hypotheses $\mathbf{(H1)}$,
$\mathbf{(H2)}$, and $\mathbf{(H3)}$ and we define:
\begin{equation}\label{exs}\phi(x,t,\varepsilon)=\int_{\Delta}\psi_\varepsilon(x,E)e^{\frac{-itE}{\varepsilon}}Q(E,\varepsilon)dE=
\sum\limits_{j=1}^{md}\phi_{j}(x,t,\varepsilon),
\end{equation}
where
\begin{equation}\label{ttt}
\phi_{j}(x,t,\varepsilon)=\int_{\Delta}c_{j}(x,E,\varepsilon)e^{\frac{-i\int_{0}^{x}k_{j}(y,E)dy}{\varepsilon}}\varphi_{j}(x,E)e^{\frac{-itE}{\varepsilon}}Q(E,\varepsilon)dE.
\end{equation}
Since the integrand is smooth and $\Delta$ is compact,
$\phi(x,t,\varepsilon)$ is an exact solution of \eqref{evol}.

We also get from the decomposition (\ref{eq:dec}) for all
$l=0,\cdots, m-1$,
\begin{eqnarray}(i\varepsilon\partial_x)^l\phi(x,t,\varepsilon)
&=&\sum_{j=1}^{md}\int_{\Delta}c_{j}(x,E,\varepsilon)
e^{\frac{-i\int_{0}^{x}k_{j}(y,E)dy}{\varepsilon}}k_j^l(x,E)\varphi_{j}(x,E)e^{\frac{-itE}{\varepsilon}}Q(E,\varepsilon)dE\\
\nonumber &\equiv&\sum_{j=1}^{md}\phi_j^{[l]}(x,t,\varepsilon),
\end{eqnarray}
with the convention
$\phi_j^{[0]}(x,t,\varepsilon)=\phi_j(x,t,\varepsilon)$. Note,
however, that in general $(i\varepsilon\partial_x)^l
\phi_j(x,t,\varepsilon)\neq \phi_j^{[l]}(x,t,\varepsilon)$.

\smallpagebreak The behavior of $\phi_{j}(x,t,\varepsilon)$ for
large $x$ can be understood under the following supplementary
assumption
\\
{\bf (GV):}
\begin{equation}\label{gr_vel}
\forall j\in\{1,\dots, md\},\quad\forall E\in\Delta,\quad
\partial_{E}k_{j}(\pm\infty,E)\neq 0.
\end{equation}
Let us note that condition ({\bf GV}) is quite natural. Indeed,
with our sign conventions, $-1/\partial_{E}k_j$ is the group
velocity of the asymptotic waves  (\ref{eq:aw}). Our condition
says that we want to describe waves with non-zero asymptotic
velocity. Moreover, {\bf (GV)} also imposes the presence of at
least one time derivative in the definition of the differential
operator ${\mathcal R}(x,
i\varepsilon\partial_t,i\varepsilon\partial_x, \delta)$.

We have
\begin{Le}\label{le:large_t}
Assume that $\mathbf{(H1)}$, $\mathbf{(H2)}$, $\mathbf{(H3)}$,
$\mathbf{(C0)}$ and $\mathbf{(GV)}$ are satisfied. Let
$$K_{+}=\sup_{E\in\Delta,\ j\in\{1,\dots,
md\}}\frac{1}{|\partial_{E}k_{j}(\pm\infty,E)|}>0$$ and
$$K_{-}=\inf_{E\in\Delta,\ j\in\{1,\dots,
md\}}\frac{1}{|\partial_{E}k_{j}(\pm\infty,E)|}>0.$$ Fix
$\alpha\in (0,1)$. Then, there exists $C_\varepsilon>0$ such that,
for $x$ large enough and for either $t=0$ or any $t\neq 0$ and $x$
satisfying:
$$|x/t|>\frac{K_{+}}{1-\alpha}\quad\textrm{or}\quad|x/t|<\frac{K_{-}}{1+\alpha},$$
we have for all $j=1,\cdots, md$:
$$\|\phi_{j}(x,t,\varepsilon)\|<\frac{C_\varepsilon}{|x|}\ \ \mbox{and} \ \
\|\phi_{j}^{[l]}(x,t,\varepsilon)\|<\frac{C_\varepsilon}{|x|},$$
where $l\in [0, \cdots, m-1]$ and  $\|\cdot\|$ is the norm in
$\C^{m}$.\\

\noindent Specializing to the $j'$th mode, there exist
$x_0^\pm(j)\in\R^\pm$ and $C_\varepsilon(j)$, independent of time,
such that for any $\beta\in(0,1)$ and any $l\in [0, \cdots, m-1]$,
if $|t|>1$ with
$\sign(t)=\pm\sign(\partial_{E}k_{j}(\pm\infty,E))$ and $\pm x\geq
\pm x_0^\pm(j)$ then
\begin{equation}\label{astx}
\|\phi_{j}^{[l]}(x,t,\varepsilon)\|<\frac{C_\varepsilon(j)}{|t|^\beta
|x|^{(1-\beta)}}.
\end{equation}
\end{Le}
\noindent{\bf Remarks:}\\
i) As direct corollaries, we get that $\phi_j^{[l]}(\cdot,
t,\varepsilon)$, and thus $(\varepsilon\partial_x)^l\phi(\cdot,
t,\varepsilon)$ belong to $L^2(\R)$, for any $t\in\mathbb R$, and
any $l=0,\cdots, m-1$. Moreover,
\begin{equation}\label{nort}
\sup_{|t|\leq 1}\|(\varepsilon\partial_x)^l\phi(\cdot,
t,\varepsilon) \|_{L^2(\mathbb R)}=O(C_\varepsilon).
\end{equation}
\\
ii) The behavior in $\varepsilon$ of $C_\varepsilon$ and
$C_\varepsilon(j)$ cannot be estimated under hypothesis {\bf (C0)}
only. However, anticipating on our eventual choice of
$Q(E,\varepsilon)$, see (\ref{howl}) below, if the energy density
satisfies
\begin{equation}\label{bdq}
\sup_{E\in\Delta \atop \varepsilon >0}(|Q(E,\varepsilon)|+
|\varepsilon\partial_E Q(E,\varepsilon)|)<\infty,
\end{equation}
 and assuming (\ref{ccb}),  then the constants
$C_\varepsilon$ and $C_\varepsilon(j)$ are actually uniform in
$\varepsilon\rightarrow 0$, as easily checked from the proof and
Lemma \ref{le:as}.\\
iii) The complicated looking second statement simply says the
following, for $0<\beta<1/2$. In the asymptotic regions where
$\phi_j^{[l]}$ is driven by the asymptotic group velocity
$-1/\partial_E k_{j}(\pm\infty,E)$, if time flows in the wrong
direction, in the sense that the wave is driven out of these
regions, then the $L^2$ norm over those regions decreases.
\\
iv) We prove this Lemma in section \ref{sec:large_t}.
 \smallpagebreak

In a scattering regime, we expect our solutions to behave as
freely propagating waves along independent modes. Let us
introduce such asymptotic  waves
$\phi(x,t,\varepsilon,\pm\infty)$:
$$\phi(x,t,\varepsilon,\pm\infty)=\sum\limits_{j=1}^{md}\phi_{j}(x,t,\varepsilon,\pm\infty),$$
with
\begin{equation}\label{eq:aw}
\phi_{j}(x,t,\varepsilon,\pm\infty)=\int_{\Delta}c_{j}
(\pm\infty,E,\varepsilon)e^{\frac{-i(k_{j}(\pm\infty,E)x+\omega_{j}
(\pm\infty,E))}{\varepsilon}}\varphi_{j}(\pm\infty,E)e^{\frac{-itE}
{\varepsilon}}Q(E,\varepsilon)dE.
\end{equation}
With respect to (\ref{ttt}), the only dependence left in the space
variable in the integrand is in the exponent. The index
$\pm\infty$ refers to the choice of asymptotic mode
$k_j(\pm\infty,E)$ and polarization $\varphi_{j}(\pm\infty,E)$
taken in the definition. Note the relation
\begin{eqnarray}\label{derass}
&
&(i\varepsilon\partial_x)^l\phi_{j}(x,t,\varepsilon,\pm\infty)=\\
\nonumber & & \int_{\Delta}c_{j}(\pm\infty,E,\varepsilon)
e^{{-i(k_{j}(\pm\infty,E)x+\omega_{j}(\pm\infty,E))}/{\varepsilon}}
k_{j}^l(\pm\infty,E)\varphi_{j}(\pm\infty,E)e^{{-itE}/
{\varepsilon}}Q(E,\varepsilon)dE \equiv\\ \nonumber & &
\phi_{j}^{[l]} (x,t,\varepsilon,\pm\infty).
\end{eqnarray}
We also remark that since
$\phi_{j}^{[l]}(x,t,\varepsilon,\pm\infty)$ are constructed as
integrals in the same way as $\phi_{j}^{[l]}(x,t,\varepsilon)$
are, only with simpler integrands, then they also satisfy the
estimates based on this structure. In particular, (\ref{astx})
holds without restriction
on the boundary of the $x$-region:\\
For any $x_0\in\R$, there exists a constant
$C^\pm_\varepsilon(j,x_0)$ such that for any $\beta\in(0,1)$ and
any $l\in [0, \cdots, m-1]$, if $|t|>1$ with
$\sign(t)=\pm\sign(\partial_{E}k_{j}(\pm\infty,E))$ and $\pm x\geq
\pm x_0^\pm(j)$, then
\begin{equation}\label{astxa}
\|\phi_{j}^{[l]}(x,t,\varepsilon,\pm\infty)\|<
\frac{C^\pm_\varepsilon(j, x_0)}{|t|^\beta |x|^{(1-\beta)}}.
\end{equation}
Again, assuming  (\ref{ccb}) and (\ref{bdq}),
$C^\pm_\varepsilon(j, x_0)$ can be chosen uniformly as
$\varepsilon\rightarrow 0$.

Finally, $\phi_{j}^{[l]}(x,0,\varepsilon,-\infty)$ determines
$\phi_{j}^{[l]}(x,0,\varepsilon,+\infty)$ by means of
(\ref{eq:diff_c}).
\\

While the waves $\phi_{j}(x,t,\varepsilon,\pm\infty)$ are not
localized in space, we expect them to be approximations of
solutions to (\ref{evolagain}) in neighborhoods of $x=\pm\infty$
only. Hence the following construction:

Let $x\mapsto \omega(x)\in [0,1]$ be a function such that
$\omega(x)=1 $ if $x\geq 1$ and $\omega(x)=0$ if $x\leq 0$. We
define asymptotic waves corresponding to
$\phi_{j}(x,t,\varepsilon,+\infty)$ for $x>1$ and to
$\phi_{j}(x,t,\varepsilon,-\infty)$ for $x<-1$ as follows:
\begin{eqnarray}\label{defas}
\phi_{j}^{[l]}(x,t,\varepsilon,a)&=& \omega(x)\phi_{j}^{[l]}
(x,t,\varepsilon,+\infty)+(1-\omega(x)) \phi_{j}^{[l]}
(x,t,\varepsilon,-\infty).\nonumber\\
\phi^{[l]}(x,t,\varepsilon,a)&=&\sum_{j=1}^{md}\phi_{j}^{[l]}(x,t,\varepsilon,a).
\end{eqnarray}

Under our hypotheses, it is easy to compute the $L^2$ norm of
these different asymptotic states by means of the rescaled Fourier
transform $\mathcal{F}_{\varepsilon}$ defined as:
\begin{equation}
\label{eq:fourier} (\mathcal{F}_{\varepsilon}\,g(\cdot))(x)\ =\
\frac{1}{\sqrt{2\pi\varepsilon}}\ \int_{\R}\
g(k)\,e^{-ikx/\varepsilon}\,dk.
\end{equation}

\begin{Le}\label{noras}
Assume $\mathbf{(H1)}$, $\mathbf{(H2)}$, $\mathbf{(H3)}$,
$\mathbf{(C0)}$ and $\mathbf{(GV)}$. Then there exists
$D_\varepsilon$ such that for all $j=1,\cdots,md$, all
$l=0,\cdots, m-1$ and all $t\in\R$
$$\|(i\varepsilon\partial_x)^l\phi_{j}(\cdot,t,\varepsilon,\pm\infty)
\|_{L^2(\mathbb R)}\leq D_\varepsilon. $$
\end{Le}

\noindent{\bf Remarks:}\\
i) As a direct corollary,
$\|(i\varepsilon\partial_x)^l\phi(\cdot,t,\varepsilon,\pm\infty)\|=O(D_\varepsilon).$
Moreover, $\|\phi_j^{[l]}(\cdot,t,\varepsilon,a)\|_{L^2(\mathbb
R)}=O(D_\varepsilon)$ and therefore
$\|\phi^{[l]}(\cdot,t,\varepsilon,a)\|_{L^2(\mathbb
R)}=O(D_\varepsilon)$.
\\
ii) Again, further assuming (\ref{bdq}) and (\ref{ccb}), we get
$D_\varepsilon=\sqrt{\varepsilon}D$, with $D$
uniform in $ \varepsilon$.\\

\demo Under  $\mathbf{(GV)}$, the reciprocal functions of
$E\mapsto k_{j}(\pm\infty,E)$ all exist on $\Delta$ and we denote
them by $k\mapsto E^\pm_j(k)$, $j=1,\cdots, md.$ Hence, using
(\ref{derass}) and a change of variables, we can write
$$
(i\varepsilon\partial_x)^l\phi_{j}(x,t,\varepsilon,\pm\infty)=
\phi_{j}^{[l]}(x,t,\varepsilon,\pm\infty)=
\sqrt{2\pi\varepsilon}(\mathcal{F}_{\varepsilon}\,
\widehat{\phi}_{j}^{[l]}(\cdot,t,\varepsilon,\pm\infty) )(x),
$$
where
$$\widehat{\phi}_{j}^{[l]}(k,t,\varepsilon,\pm\infty)
=c_{j}(\pm\infty,E^\pm_j(k),\varepsilon)
e^{{-i\omega_{j}(\pm\infty,E^\pm_j(k))}/{\varepsilon}}
{k}^l\varphi_{j}(\pm\infty,E^\pm_j(k))e^{{-itE^\pm_j(k)}/
{\varepsilon}}Q(E^\pm_j(k),\varepsilon)\partial_kE^\pm_j(k). $$ By
Plancherel formula,
$$\|(i\varepsilon\partial_x)^l\phi_{j}(\cdot,t,\varepsilon,\pm\infty)
\|_{L^2(\mathbb R)}=\sqrt{2\pi\varepsilon}
\|\widehat{\phi}_{j}^{[l]}(\cdot,t,\varepsilon,\pm\infty)
\|_{L^2(\mathbb R)}\equiv D_\varepsilon ,$$
where $D_\varepsilon$ is uniform in $t\in\R$.\\

Finally, as expected, we show that the exact solutions (\ref{exs})
behave more and more like the corresponding free asymptotic waves
(\ref{defas}), in $L^2$ norm, as time gets large. Furthermore,
 we show that (\ref{exs}) cannot get trapped on a compact
set of $\R$ as time goes to infinity, since its $L^2$ norm
vanishes  for $|t|\rightarrow \infty$ on such sets:
\begin{Pro}
\label{pro:asl_2} Assume that $\mathbf{(H1)}$, $\mathbf{(H2)}$,
$\mathbf{(H3)}$, $\mathbf{(GV)}$ and $\mathbf{(C0)}$ are
satisfied. Then, there exists $C_\varepsilon>0$ such that we have
for any  $|t|>0$,
 $\forall j\in\{1,\dots,md\}$ and $\forall l\in\{0,\dots,m-1\}$:
$$\|\phi_j^{[l]}(\cdot,t,\varepsilon)-
\phi_j^{[l]}(\cdot,t,\varepsilon,a)\|_{L^2(\mathbb
R)}<\frac{C_\varepsilon}{|t|}.$$ Moreover, for any bounded
interval $I\in\R$,
$$\|\phi_j^{[l]}(\cdot,t,\varepsilon)\|_{L^2(I)}<
\frac{\tilde{C}_\varepsilon}{|t|}, $$ for some
$\tilde{C}_\varepsilon$ depending on $I$.
\end{Pro}

\noindent {\bf Remarks:}\\
i)  As a direct corollary, $$\|(i\varepsilon\partial_{x})^{l}
\phi(\cdot,t,\varepsilon)-
\phi^{[l]}(\cdot,t,\varepsilon,a)\|_{L^2(\mathbb
R)}<\frac{C_\varepsilon}{|t|}.$$ ii)  Further assuming (\ref{bdq})
and (\ref{ccb}), we can take $C_\varepsilon=C$, and
$\tilde{C}_\varepsilon= \tilde{C}$ uniformly as
$\varepsilon\rightarrow 0$, see the proof.\\
iii) The estimate is independent of the signs of $t$ and of the
asymptotic group velocities, because the definition of
$\phi^{[l]}(\cdot,t,\varepsilon,a)$ takes into account the
asymptotic waves travelling in both asymptotic regions. See the
example below for an illustration.  \\
iv) We  prove this Proposition in \ref{sec:asl_2}.\\

In order to have a better understanding of the localization
properties for large times of the asymptotic approximation
$\phi_j^{[l]}(\cdot,t,\varepsilon,a)$, we need to look at the
signs of the group velocities $-/\partial_E k_j(\pm\infty, E)$ of
its components (\ref{eq:aw}). Different cases occur that we list
below.

\begin{Cor} Assume $\mathbf{(H1)}$, $\mathbf{(H2)}$,
$\mathbf{(H3)}$, $\mathbf{(GV)}$ and $\mathbf{(C0)}$ are
satisfied. Then there exists a constant $H_\varepsilon$ such that
and for any $0<\beta <1/2$ and $|t|\geq 1$,
\begin{eqnarray}
&&\left\{\ \partial_E k_j(-\infty, E)\partial_E k_j(+\infty, E)<0\
\ \mbox{and} \
\ t\partial_E k_j(+\infty, E)<0\ \right\}\nonumber \\
&& \qquad\qquad\qquad\qquad \Rightarrow
\|\phi_j^{[l]}(\cdot,t,\varepsilon,a)-
(\phi_j^{[l]}(\cdot,t,\varepsilon,-\infty)+
\phi_j^{[l]}(\cdot,t,\varepsilon,+\infty))\|_{L^2(\mathbb R)}
\leq\frac{H_\varepsilon}{|t|^{\beta}},\nonumber\\
&&\left\{\ \partial_E k_j(-\infty, E)\partial_E k_j(+\infty, E)>0\
\ \mbox{and} \
\ t\partial_E k_j(+\infty, E)>0\ \right\}\nonumber \\
&& \qquad\qquad\qquad\qquad \Rightarrow
\|\phi_j^{[l]}(\cdot,t,\varepsilon,a)-
\phi_j^{[l]}(\cdot,t,\varepsilon,-\infty)\|_{L^2(\mathbb R)}
\leq\frac{H_\varepsilon}{|t|^{\beta}},\nonumber\\
&&\left\{\ \partial_E k_j(-\infty, E)\partial_E k_j(+\infty, E)>0\
\ \mbox{and} \
\ t\partial_E k_j(+\infty, E)<0\ \right\}\nonumber \\
&& \qquad\qquad\qquad\qquad \Rightarrow
\|\phi_j^{[l]}(\cdot,t,\varepsilon,a)-
\phi_j^{[l]}(\cdot,t,\varepsilon,+\infty)\|_{L^2(\mathbb R)}
\leq\frac{H_\varepsilon}{|t|^{\beta}},\nonumber\\
&&\left\{\ \partial_E k_j(-\infty, E)\partial_E k_j(+\infty, E)<0\
\ \mbox{and} \
\ t\partial_E k_j(+\infty, E)>0\ \right\}\nonumber \\
&& \qquad\qquad\qquad\qquad \Rightarrow
\|\phi_j^{[l]}(\cdot,t,\varepsilon,a) \|_{L^2(\mathbb R)}
\leq\frac{H_\varepsilon}{|t|^{\beta}},\nonumber
\end{eqnarray}

\end{Cor}
\demo Just make use of the definition (\ref{defas}),  of
(\ref{astxa}) and
of the support properties of $\omega$.\\

\noindent{\bf Remark:} Again, if (\ref{bdq}) and (\ref{ccb}) are
true, the constant $H_{\varepsilon}$ is uniform
in $\varepsilon$.\\

Another consequence of  Lemmas \ref{le:large_t} and \ref{noras},
equation (\ref{nort}) and Proposition \ref{pro:asl_2} is the
following estimate
\begin{Cor} Assume $\mathbf{(H1)}$, $\mathbf{(H2)}$,
$\mathbf{(H3)}$ and $\mathbf{(C0)}$ are satisfied. Then, there
exists $F_\varepsilon>0$  such that for all $l=0,\cdots, m-1$,
$$\sup_{t\in\mathbb R}\|(\varepsilon\partial_x)^l \phi(x,t,\varepsilon)
\|_{L^2(\mathbb R)} \leq F_\varepsilon.$$ If, furthermore,
(\ref{bdq}) and (\ref{ccb}) are true, $F_\varepsilon$ can be
chosen as $F$, uniform in $\varepsilon\rightarrow 0$.
\end{Cor}
Hence, if the $L^2$ norm is not conserved under the time evolution
(\ref{evol}), it remains uniformly bounded in time. Moreover, it
is also uniformly bounded in $\varepsilon$, for the type of energy
densities that we will use below, see (\ref{howl}),  with
(\ref{bdq}) and (\ref{ccb}). Hence, in that case, the $L^2$ norm
of our solutions at any time is proportional to that they had at
any initial time $t_0$:
\begin{equation}\label{diffinit}\|\phi(\cdot,t,\varepsilon)\|_{L^2(\mathbb R)}\leq F
\|\phi(\cdot,t_0,\varepsilon)\|_{L^2(\mathbb R)}.
\end{equation}

Let us illustrate some of the notions of this Section by means of
an explicitly solvable example. Consider the following scalar
linear PDE:
\begin{equation}\label{th}
( \tanh
(x)i\epsilon\partial_t-i\epsilon\partial_x)\phi(x,t,\epsilon)=0.
\end{equation}
The corresponding dispersion relation yields $k(x,E)=E\tanh (x)$
as unique mode, which satisfies {\bf (GV)}. As the equation is
$\varepsilon$-independent, we take $\varepsilon=1$. The general
solution reads $\phi(x,t,1)=f(t+\ln(\cosh(x))),$ where $f$ is any
regular function. To have a solution obtained by means of a
superposition of generalized eigenvectors $e^{-iE\int_0^x\tanh
(y)dy}$ according to some compactly supported energy density, we
must have
$$
f(t+\ln(\cosh(x)))= \int_{\Delta}e^{-iE\int_0^x\tanh (y)dy}
e^{-iEt}Q(E,1)dE=\int_{\Delta} e^{-iE(t+\ln(\cosh(x)))}Q(E,1)dE.
$$
Hence, $f=\sqrt{2\pi}(\mathcal{F}_1Q(\cdot,1))$ and is therefore
$L^2$, analytic and goes to zero at infinity.

That the $L^2$ norm is not conserved in general under our
hypotheses is now easily seen: Since $0\leq \ln(\cosh(x))$ is even
and behave as $|x|$ for $x$ large, one  checks that we have
$\lim_{t\rightarrow -\infty}\|\phi(\cdot,t,1)\|_{L^2(\mathbb R)}
=O(\|f(\cdot)\|_{L^2(\mathbb R)})>0$, whereas, $\lim_{t\rightarrow
+\infty}\|\phi(\cdot,t,1)\|_{L^2(\mathbb R)}=0$.

Let us investigate the asymptotic waves corresponding to
(\ref{th}). Using $k(\pm\infty,E)=\pm E$, we find
$$\phi(x,t,1,\pm\infty)= \int_{\Delta}e^{ \mp i x E}
e^{-iEt}Q(E,1)dE=\sqrt{2\pi}(\mathcal{F}_1Q(\cdot,1))(t\pm x).$$
Hence,
$$\phi(x,t,1,a)=\omega(x)\sqrt{2\pi}(\mathcal{F}_1Q(\cdot,1))
(t+ x)+(1-\omega(x))\sqrt{2\pi}(\mathcal{F}_1Q(\cdot,1))(t- x), $$
which, as $t\rightarrow -\infty$, is significant at both large and
positive values of $x$ and large and negative values of $x$.
Accordingly, for $t\simeq -\infty$, $f(t+\ln(\cosh(x)))$ is
significant at values of $\ln(\cosh(x))\simeq |x|\simeq |t|$, i.e.
for $x\simeq \pm |t|$. The picture is that of two bumps at plus
and minus infinity in space that travel towards one another with
unit velocity and disappear as they collide. This is correctly
captured by the approximation $\phi(x,t,1,a)$, for large times.

\section{Semi-classical transitions asymptotics}
\label{sec:transas}
\subsection{The transition integral}
We assume here we are in an avoided crossing situation, and we do
not explicit the dependence in the variable $\delta>0$ in the
notation. We have obtained the asymptotics of the scattering
matrix $S(E,\varepsilon)$ in Section \ref{sec:wkb}. We now compute
the small $\varepsilon$-asymptotics of the integrals that describe
the asymptotic states $\phi_j(x,t,\varepsilon,\pm\infty)$ given by
(\ref{eq:aw}) as $|t|\rightarrow\infty$, for the different
channels.\smallpagebreak We assume that $j$ is such that
(\ref{inco}) holds and let $n=\pi(j)$ be given by (\ref{eq:perm}).
\smallpagebreak We choose our energy density $Q(E,\varepsilon)$ to
be more and more sharply peaked near a specific value $E_0\in
\Delta \setminus \partial \Delta$ as $\varepsilon\rightarrow 0$.
As a result, we obtain semiclassical wavepackets  that are well
localized in phase space. This is a physically reasonable choice
that allows for a complete semiclassical treatment.

More precisely we consider, \begin{equation}\label{howl}
Q(E,\varepsilon)\ =\ e^{-\,G(E)/\varepsilon}\
e^{-\,i\,J(E)/\varepsilon}\ P(E,\varepsilon), \end{equation} where
\begin{description}
\item[(C1)] The real-valued function $G\ge 0$ is in
$C^3(\Delta)$, is independent of $\delta$ and has a unique
non-degenerate absolute minimum value of $0$ at $E_0$ in the
interior of $\Delta$. This implies that
$$
G(E)\ =\ g\,(E-E_0)^2/2\ +\ O(E-E_0)^3,\quad\mbox{ where }\quad
g>0.
$$
\item[(C2)] The real-valued function $J$ is in $C^3(\Delta)$.
\item[(C3)] The complex-valued function $P(E,\varepsilon)$ is in
$C^1(\Delta)$ and satisfies \begin{equation} \sup_{E\in\Delta
\atop \varepsilon \geq 0}\ \left|\,\frac{\partial^n}{\partial
E^n}\,P(E,\varepsilon)\,\right|\ \leq\ C_n,\qquad\mbox{for}\qquad
n=0,\,1.\label{eq:condd} \end{equation}
\end{description}
\noindent {\bf Remarks:}\\
i) Typical interesting choices of $Q$ are $G\,=\,g\,(E-E_0)^2$,
$J=0$, and $P$ an $\varepsilon$-dependent multiple (the equation
(\ref{evol}) is linear) of a smooth function with at most
polynomial growth in $(E-E_0)/\varepsilon$.\\
ii) We want to emphasize the fact that a Gaussian energy density
does not give rise in general to a Gaussian solution. See the
discussion in the Introduction and Section $6$ of \cite{JH:04}.

\smallpagebreak The leading inter-modes transitions are described
by the asymptotics of those coefficients $c_l(\pm\infty,
E,\varepsilon)$ that satisfy
\begin{eqnarray}\label{eq:choice}
c_k(-\infty,E,\varepsilon)&=&\delta_{j,k}\\
\label{eq:choice2}
c_n(+\infty,E,\varepsilon)&=&e^{-i\theta_j(\zeta,E)}\
e^{i\int_{\zeta}k_j(z,E)dz/\varepsilon} \ (1+O_E(\varepsilon)),
\end{eqnarray} where $n=\pi(j)=j\pm 1$. We recall that the error term
$O_E(\varepsilon)$ depends analytically on the energy $E$ in a
neighborhood of the compact set $\Delta$. We have already noted in
the comments after Theorem \ref{th:PERCO} that the term
$O_E(\varepsilon)$ satisfies (\ref{eq:condd}).

\begin{Th}\label{th:astrans}
Assume {\bf (H1)} to {\bf (H3)}, {\bf (AC)} and {\bf (GV)}. Let
$Q(\cdot, \epsilon)$ be the energy density supported on the
interval $\Delta$ defined in (\ref{howl}) which satisfies {\bf
(C1)}, {\bf (C2)}, and {\bf (C3)}. Let $\phi(x,t,\varepsilon)$ be
a solution of equation (\ref{evolagain}) of the form (\ref{exs}).
Assume $\partial_Ek_j(-\infty,E)<0$ on $\Delta$, for some $j$ and
suppose that the solution is characterized in the past by
$$ \lim\limits_{t\rightarrow
-\infty}\|\phi(\cdot,t,\varepsilon)-\phi(\cdot,t,\varepsilon,a)
\|_{L^2(\mathbb R)}=0,$$ where, as $t\rightarrow -\infty$,
$$\omega(x)\phi(x,t,\varepsilon,a)=
\int_{\Delta}Q(E,\varepsilon)e^{-itE/\varepsilon}e^{-i(xk_{j}(-\infty,E)
+\omega_{j}(-\infty,E))/\varepsilon}\varphi_{j}(-\infty,E)dE+O(1/|t|^\beta).$$
 Let $n=\pi(j)$ be given by (\ref{eq:perm}), and
let\begin{eqnarray}\label{eq:al}
\alpha(E)&=&G(E)\,+\, \im (\int_{\zeta}\,k_j(z,E)\,dz),\\
\label{eq:ka} \kappa(E)&=&J(E)\,-\,
\re(\int_{\zeta}\,k_j(z,E)\,dz)\,+\,\omega_n(+\infty,E).
\end{eqnarray} Assume $E^*$ is the unique absolute minimum of
$\alpha(\cdot)$ in {\em Int}\,$\Delta$ and define
$k^{*}=k_{n}(+\infty,E^{*})$. Let $k\mapsto E_n^\pm(k)$ be the
inverse function of $E\mapsto k_{n}(\pm\infty, E)$ on
$\Delta$.\smallpagebreak Then, there exist $\delta_0>0$, $p>0$
arbitrarily close to $5/4$, and a function
$\varepsilon_0:(0,\delta_0)\rightarrow \R^+$, such that for all
$0<\beta<1/2$, $\delta <\delta_0$, and
$\varepsilon<\varepsilon_0(\delta)$, the following asymptotics
hold as $t\rightarrow -\sign(\partial_k E_n^+(k^*))\infty$, in
$L^2(\mathbb R)$ norm:
\begin{eqnarray}
& &\phi_{n}(x,t,\varepsilon)=
\sqrt{2\pi\varepsilon}P(E^{*},\varepsilon)
e^{-\alpha(E^{*})/\varepsilon}e^{-i\kappa(E^{*})/\varepsilon}\varphi_n(+\infty,E^{*})
e^{-\,i\,\theta_j(\zeta,E^{*})}\partial_{k}E_{n}^+(k^{*})\times \\
 \nonumber & &\qquad\qquad\qquad\times\mathcal{F}_{\varepsilon}(e^{-itE^+_{n}(\cdot)/\varepsilon}e^{-\Lambda(\cdot)/\varepsilon}\chi_{k_{n}(+\infty,\Delta)})
 + O(e^{-\alpha(E^*)/\varepsilon}\varepsilon^{p})
+O\left(1/|t|^{\beta}\right),
\end{eqnarray}
where
$$\Lambda(k)=\frac{\lambda_{2}}{2}(k-k^{*})^{2}+i\lambda_{1}(k-k^{*}),\
\mbox{with} \
\lambda_{1}=\partial_{k}E_{n}^+(k^{*})\kappa'(E^{*}),$$
$$\lambda_{2}=[\partial_{k}E_{n}^+(k^{*})]^{2}\alpha"(E^{*})+i\left[\kappa"(E^{*})[\partial_{k}E_{n}^+(k^{*})]^{2}+\kappa'(E^{*})\partial^{2}_{k}E_{n}^+(k^{*})\right],$$
and $\chi_{k_{n}(+\infty,\Delta)}$ is the characteristic function
of the set
$k_{n}(+\infty,\Delta)$.\\
Moreover, if $t\rightarrow \sign(\partial_k E_n^+(k^*))\infty$,
then $\|\phi_{n}(x,t,\varepsilon)\|_{L^2(\mathbb
R)}=O\left(1/|t|^{\beta}\right)$.
\end{Th}
\noindent{\bf Remarks:} \\
0) The first error term is uniform in $t$ whereas the second error
term is
uniform in $\varepsilon$.\\
i) The same result holds for $\phi$, and $\phi_n(+\infty)$,
replaced by $(i\varepsilon\partial)^l \phi$, and
$(i\varepsilon\partial)^l \phi_n(+\infty)$ respectively, with
$\{l=0,\cdots, m-1\}$,
at the expense of a multiplication of the prefactor by $k_n(+\infty, E^*)^l$.  \\
ii) As will be made explicit in  Section \ref{spti} below, the
$L^2$ norm of the leading term expressed as a Fourier transform is
positive, of order ${\varepsilon^{1/4}}$, and independent of time.
The leading term hence becomes meaningful for times $t$ that are
of order $|t|\simeq e^{c/\varepsilon}$, for some $c>0$  at least.
We get control over this time scale far beyond the Ehrenfest or
Heisenberg times of Quantum
semiclassical analysis thanks to our scattering setup.  \\
iii) The leading term clearly satisfies the asymptotic PDE (\ref{asedp}).\\
iv) The energy $E^*$ depends explicitly on the properties of the
involved modes and on the energy density $Q(E,\varepsilon)$ as well.\\
v) The space-time localization properties of the leading term are
further discussed in
Section \ref{spti}.\\
vi) Also, as mentioned earlier, we can specify the coefficients
$c_l$ at $x=+\infty$ instead.\\
vii) The proof of the Theorem is given in the last Section of the paper.\\

Let us finally discuss our hypotheses and interpret our result.
The condition on the sign of $\partial_E(k_j(-\infty,E))$ says
that the group velocity of $\phi_j(x,t,\varepsilon,-\infty)$ is
positive, so that $\phi(x,t,\varepsilon,a)$ is non trivial as
$t\rightarrow -\infty$ for negative $x$'s and describes an ingoing
wave. If the asymptotic group velocity of the mode $k_n(x,E)$ is
positive as $x\rightarrow +\infty$, our results describes an
outgoing  {\em transmitted wave} for large positive times, as
discussed in the introduction. If the asymptotic group velocity
$-\partial_{k}E_{n}^+(k)$ is negative, we describe another ingoing
wave along mode $n$, for large negative times and large positive
$x$'s, arising during the evolution, which, as time goes to
$+\infty$, goes to zero. Note also that if the asymptotic group
velocity of mode $k_j(x,E)$ at $x=+\infty$ is positive, then an
order one wave, in the sense that
$c_j(+\infty,E,\varepsilon)=1+O(\varepsilon)$, propagates along
positive $x$'s for positive times. If the asymptotic group
velocity of mode $k_j(x,E)$ at $x=+\infty$ is negative, there is
no wave propagating along positive $x$'s to the right, for large
positive times, but another ingoing wave from large positive $x$'s
and large negative times.

Therefore, in case $-\partial_{k}E_{n}^+(k)$ is negative and both
$-\partial_{k}E_{j}^-(k)$ and $-\partial_{k}E_{j}^+(k)$ are
positive, running the evolution {\em backwards in time}, we have
an ingoing wave (of order one in the sense above) on mode $j$, for
$x\rightarrow +\infty$ and $t\rightarrow +\infty$, and, as
$t\rightarrow -\infty$, we have an  outgoing wave on mode $j$, for
$x\rightarrow -\infty$, and another exponentially small outgoing
wave on mode $n$, for $x\rightarrow +\infty$, whose asymptotics is
determined by our Theorem. Hence, we describe the asymptotics of a
{\em reflected wave} in mode $n$. Note that reflected waves on
other modes may be present as well. In any case, they are
exponentially small.

Finally, in case  $-\partial_{k}E_{n}^+(k)$ and
$-\partial_{k}E_{j}^+(k)$ are both negative, we describe a
scattering process in which we have ingoing solutions on the modes
$j$ and $n$, that all disappear as time goes to $+\infty$, in a
similar way as what happens in the illustration ending the
previous Section.

\subsection{Perturbative results in $\delta$}
We assume that $\mathbf{(H4)}$ is also satisfied and restore back
$\delta$ in the notation. We have the following sharper result
concerning the behavior as $\delta\rightarrow 0$ of the quantities
involved in the description of the asymptotic wave:
\begin{Pro}
Further assuming $\mathbf{(H4)}$, we have the following as
$\delta\rightarrow 0 $, for $E\in \Delta$:
$$\im\int_{\zeta}(k_{i}-k_{j})(z,E,\delta)\ dz=D(E)\delta^{2}+O(\delta^{3}),$$ with
$D(E)=\frac{\pi}{4} \frac{a^{2}(E)b^{2}(E)-c^{2}(E)}{a^{3}(E)}.$
\smallpagebreak This implies that $(E,\delta)\mapsto
\im\int_{\zeta}(k_{i}-k_{j})(z,E,\delta)dz$ is a positive
function.\smallpagebreak Let
$\alpha(E,\delta)=G(E)+\im\int_{\zeta}(k_{i}-k_{j})(E,\delta)$.
There exists $E^{*}(\delta)$ such that
$$\partial_{E}\alpha(E^{*}(\delta),\delta)=0.$$ It satisfies:
$$E^{*}(\delta)=E_{0}-\frac{D'(E_{0})}{g}\delta^{2}+O(\delta^{3}).$$
\end{Pro}
The results above hold provided one knows $E^{*}(\delta)$ is the
unique absolute minimum of $\alpha$ in the set $\Delta$, which is
generically true. Again, if there are several minima, one simply
adds the corresponding contributions. Note also that if the
constant $g$ characteristic of the energy density is of order
$\delta^2$, the difference $E_0-E^*$ is of order one as $\delta$
shrinks to zero. This corresponds to a ``wide'' energy density of
width $\varepsilon/\delta^2$ around $E_0$. This result is a
straightforward consequence of the Implicit Function Theorem, the
proof of which we omit.

\subsection{Explicit computation in case $E_{n}^+(k)$ is quadratic}
In this paragraph, we assume that $k\mapsto E_{n}^+(k)$ is
quadratic:
\begin{equation}
\label{eq:quad} \forall k\in
k_{n}(\Delta,+\infty),\quad\partial_{k}^{3}E_{n}^+(k)= 0.
\end{equation}
This is true for all modes in the study of the Born-Oppenheimer
approximation, see \cite{JH:04}. This situation allows for an
explicit determination of the leading term in the asymptotic wave.
We also assume that the function $\alpha$ has a unique absolute
minimum $E^{*}(\delta)$. For sufficiently small $\delta$, this
minimum is non degenerate and satisfies
$E^{*}(\delta)\in\textrm{Int}\Delta$. \smallpagebreak The
following result is proven in Section \ref{sec:techn}:
\begin{Le}
\label{le:quad} Assume that $k\mapsto E_{n}^{+}(k)$ is quadratic
and that $\alpha$ has a unique absolute minimum
$E^{*}(\delta)\in\textrm{Int}\Delta$. There exists $p\in]3/4,5/4[$
such that, as $\varepsilon\rightarrow 0$:
\begin{eqnarray}
&&\phi_{n}(x,t,\varepsilon,+\infty)=
e^{-\alpha(E^{*})/\varepsilon}e^{-i\kappa(E^{*})/\varepsilon}\varphi_n(+\infty,E^{*})e^{-\,i\,\theta_j(\zeta,E^{*})}
P(E^{*},\varepsilon)\partial_{k}E_n^+(k^{*})\times \\
 \nonumber & &\qquad\qquad\qquad\times
 \frac{\sqrt{2\pi\varepsilon}e^{-i(k^{*}x+tE^{*})/\varepsilon}}{[\lambda_{2}+i\partial_{k}^{2}E_{n}^{+}(k^{*})t]^{1/2}}
e^{-\frac{(\lambda_{1}+\partial_{k}E_{n}^{+}(k^{*})t+x)^{2}}{2\varepsilon(\lambda_{2}+i\partial_{k}^{2}E_n^+(k^{*})t)}}
 + O(e^{-\alpha(E^*)/\varepsilon}\varepsilon^{p}),
\end{eqnarray}
\end{Le}
\noindent {\bf Remarks:}\\
i) The leading term in that case is a freely propagating Gaussian,
i.e. an exact solution to
$$i\varepsilon \partial_t g(x,t,\varepsilon)=
\left(E^*+\partial_kE_n^+(k^{*})(i\varepsilon\partial_x-k^*)+
\frac{\partial_{k}^{2}E_n^+(k^{*})}{2}(i\varepsilon\partial_x-k^*)^2\right)g(x,t,\varepsilon),$$
centered at $x_c(t)=-\partial_kE_{n}^{+}(k^{*})t-\lambda_1$, of
width
$\sqrt{\varepsilon t}$ and of $L^2$ norm of order $\varepsilon^{3/4}$. \\
ii) In the general case,  the error terms involved in the course
of the computation are not uniform in time, which prevents us to
get such an explicit form for the asymptotic wave. Nevertheless,
we show in the next Section that we can get a fairly accurate
description of such asymptotic waves, for large times and small
$\varepsilon$.

\section{Space-time properties of the asymptotic waves}\label{spti}

As seen above, the interpretation of our results makes use of  the
space-time properties the different asymptotic waves
$\phi_{j}(x,t,\varepsilon,\pm\infty)$ in terms of which the
time-dependent scattering processes are expressed. The present
Section is devoted to a thorough description of  the space-time
properties  of the leading term of these waves  as
$\varepsilon\rightarrow 0$ and $|t|\rightarrow \infty$.
\\

We first note that Theorem \ref{th:astrans} also holds for the
wave $\phi_j(x,t,\varepsilon, \pm\infty)$, which characterized by
the asymptotics $c_j(\pm\infty, E, \varepsilon)
=1+O_E(\varepsilon)$. It suffices to replace the index $n$ by $j$,
the values $E^*$ and $k^*$ by $E_0$ and $k_0$, and to set
$\alpha(E)=G(E)$ and $\theta_j(\zeta, E)\equiv 0$. Note in
particular, that $\alpha(E_0)=0$, as it should be.

Therefore, the space-time properties of the asymptotic waves along
modes $j$ and $\pi(j)=n$ are encoded in the Fourier transform
\begin{equation}
\label{eq:aswa} \mathcal{F}_{\varepsilon}\,
(e^{-itE^\sigma_{l}(\cdot)/\varepsilon}e^{-\Lambda(\cdot)/\varepsilon}\chi_{k_{l}(+\infty,\Delta)})(x)\
=\ \frac{1}{\sqrt{2\pi\varepsilon}}\ \int_{{k_l(\sigma,\Delta)}}\
\,e^{-i(kx+tE_l^\sigma(k))/\varepsilon}e^{-\Lambda(k)/\varepsilon}\,dk,
\end{equation}
where the index $l$ stands for $j$ or $n$, and $\sigma$ for $+$ or
$-$. We will also denote $k^*$ or $k_0$, respectively $E^*$ or
$E_0$, depending on the context, by $\tilde{k}$, respectively
$\tilde{E}$. We can make use of the positivity of the real part of
the function $\Lambda(k)$ and of Parseval's formula to regularize
and localize the integrand as follows. Let $\eta\in
C_0^\infty(\R)$ with support in $[-1,1]$ and $\eta(k)\equiv 1$  in
a neighborhood of $k=0$. Set
$\eta_\varepsilon(k):=\eta((k-\tilde{k})/\varepsilon^\tau)$, with
$0<\tau<1/2$. Then, if $k\not\in \mbox{supp}(\eta_\varepsilon)$,
$|e^{-\Lambda(k)/\varepsilon}|=O(\varepsilon^\infty)$. Therefore,
we have in $L^2$ norm,
\begin{eqnarray}\label{eq:statmet}
 \mathcal{F}_{\varepsilon}\,
(e^{-itE^\sigma_{l}(\cdot)/\varepsilon}e^{-\Lambda(\cdot)/\varepsilon}
\chi_{k_{l}(+\infty,\Delta)})(x)\
&=&\frac{1}{\sqrt{2\pi\varepsilon}}\ \int_{\mathbb R}\
\,e^{-i(kx+tE_l^\sigma(k))/\varepsilon}e^{-\Lambda(k)/\varepsilon}\eta_\varepsilon(k)\,dk
+O(\varepsilon^\infty)\nonumber\\
&=&  \mathcal{F}_{\varepsilon}\,
(e^{-itE^\sigma_{l}(\cdot)/\varepsilon}e^{-\Lambda(\cdot)/\varepsilon}\eta_\varepsilon(\cdot))(x)
+O(\varepsilon^\infty),
\end{eqnarray}
where the error term is uniform in $t$. Note also that by Parseval
again,
\begin{eqnarray}\label{eq:l2}
\left\|\mathcal{F}_{\varepsilon}\,
(e^{-itE^\sigma_{l}(\cdot)/\varepsilon}e^{-\Lambda(\cdot)/\varepsilon}\eta_\varepsilon(\cdot))
\right\|^2_{L^2(\mathbb R_x)}&=&\sqrt{\varepsilon}\int_{\mathbb
R}\ e^{-\re \lambda_2 z^2}\ dz
+O(\varepsilon^\infty)\nonumber\\
&=&\sqrt{\frac{\varepsilon 2\pi}{
[\partial_{k}E_{l}^\sigma(\tilde{k})]^{2}\alpha"(\tilde{E})}}+O(\varepsilon^\infty),
\end{eqnarray}
uniform in $t$. Hence, the $L^2$ norm of the asymptotic state in
Theorem \ref{th:astrans} is positive, independent of time and of
order $\varepsilon^{3/4}$.

Now, as $k_l(\sigma\infty,\cdot)$ is analytic in $E\in\Delta$, the
same is true for the inverse function $E_l^\sigma(\cdot)$ in $k\in
k_{l}(+\infty,\Delta)$. Moreover,
$e^{-\Lambda(\cdot)/\varepsilon}\eta_\varepsilon(\cdot)$ is in
$C_0^\infty$, so that we can apply stationary phase methods to
describe the large $t$ and $x$ behavior of (\ref{eq:statmet}).
\begin{Pro} \label{stph} Let $\eta_\varepsilon$ be as above and   $1>\alpha>1/2$ and
assume $\partial_k E^\sigma_{l}(\tilde{k})\neq 0$. Define for all
$|t|\geq 1$,
$$C_t(\varepsilon)=\cup_{|k-\tilde{k}|\leq \varepsilon^\tau}
\left\{x\in\mathbb R\, | \, |x+ \partial_kE^\sigma_{l}(k)t|\leq
|t|^\alpha \right\}$$ Then, there exist $\varepsilon_0>0$ and
$c(n)>0$, such that for all $\varepsilon<\varepsilon_0$, all
$n\in\mathbb N$ and all $|t|\geq 1/ \varepsilon^{1/(1-\alpha)}$,
\begin{eqnarray}
\left\|\frac{1}{\sqrt{2\pi\varepsilon}}\ \int_{\mathbb R}\
\,e^{-i(kx+tE_l^\sigma(k))/\varepsilon}e^{-\Lambda(k)/\varepsilon}
\eta_\varepsilon(k)\,dk\right\|_{L^2(\mathbb R\setminus
C_t(\varepsilon))} &\leq&
{c(n)}\frac{\varepsilon^{1/2+\tau}}{|t|^{3\alpha/2 -1}}\left(
\frac{\varepsilon}{|t|^{2\alpha-1}}\right)^n\\ \nonumber
&=&O\left(\left(\frac{\varepsilon}{|t|^{2\alpha-1}}\right)^\infty\right).
\end{eqnarray}
\end{Pro}
\noindent {\bf Remarks:}\\
i)  The Proposition says, essentially, that the whole $L^2$ mass
of the asymptotic wave in Theorem \ref{th:astrans}  is located at
time $t$ in (a slightly larger) neighborhood of size $\sqrt{|t|}$
of the point propagating with the group velocity $-\partial_k
E^\sigma_{l}(\tilde{k})$, up to arbitrarily
small corrections as $\varepsilon/|t|^{2\alpha-1}\rightarrow 0$.\\
ii) The Proposition actually also holds if $\varepsilon=1$, if
one is not interested in the small  $\varepsilon$ behavior.\\
iii) The condition  $|t|\geq 1/\varepsilon^{1/(1-\alpha)}$
actually represents no restriction in our case, since we need to
work with exponentially large times in $\varepsilon$, in order to
have a meaningful
leading order term in Theorem \ref{th:astrans}.\\
iv) The proof is given in the last Section.\\

While we don't need to assume anything on the direction of
propagation of the involved waves for Theorem \ref{th:astrans} to
hold, it's usefulness in describing time-dependent scattering
processes is revealed by the above interpretation based on these
directions of propagation.

\section{Technicalities}
\label{sec:techn}
\subsection{Proof of Lemma \ref{le:ana}}

 We first prove that if $\Delta$ is small enough, $p$ is independent of $E$.
Fix $E_{0}\in\Delta$ and $x_{0}$ such that $(k_{i}-k_{j})(x_{0},
E_{0})=0$. By hypothesis $k_{i}-k_{j}$ is continuous, then, by
Cauchy formula, $\partial_{x}(k_{i}-k_{j})$ is continuous and
$\partial_{x}(k_{i}-k_{j})\neq 0$ in a neighborhood of
$(x_{0},E_{0})$. By local inversion, the set $\{E\in\Delta\ ;\
p(E)=p(E_{0})\}$ is open. Thus, for any $E\in\Delta$, $p(E)$ is
constant.\smallpagebreak By linear perturbation theory, there
exists $Y>0$ such that there are no non real crossings in
$\rho_{Y}$, for any $E\in\Delta$. \smallpagebreak We have the
following result:
\begin{Le}
\label{le:ana2}
\begin{enumerate} Fix $i\neq j$. The functions $k_{i}$ and $k_{j}$ have the following properties
\item The function $(z,E)\mapsto k_{i}(z,E,0)+k_{j}(z,E,0)$ is analytic on
$\rho_{Y}\times\Delta$.
\item The function $(z,E)\mapsto (k_{i}(z,E,0)-k_{j}(z,E,0))^{2}$ is analytic on
$\rho_{Y}\times\Delta$.
\end{enumerate}
\end{Le}
\demo  According to \cite{Ka}, we know that we only have to check
the analyticity of $k_{i}+k_{j}$ and $(k_{i}-k_{j})^{2}$ in a
neighborhood of a crossing, actually a branch point
$(x_{0},E_{0})$. Let $P(z,E)$ be the $2$-dimensional projector on
the $\lambda$-group corresponding to the eigenvalues
$k_{i}(z,E,0)$ and $k_{j}(z,E,0)$. Let $\Gamma$ be a small a
closed path in $\mathbb C$ surrounding $k_i(x_{0},E_{0})$. For
$(z,E)$ in a neighborhood of $(x_{0},E_{0})$, we can write :
$$P(z,E)=\frac{1}{2i\pi}\int_{\Gamma}(H(z,E,0)-\lambda)^{-1}d\lambda.$$
As $H$ and thus its resolvent on $\Gamma$ are analytic in $(z,E)$,
this implies that $(z,E)\mapsto P(z,E)$ is analytic in a
neighborhood of $(x_{0},E_{0})$. We consider
$\{\varphi_{1}(x_{0},E_{0}),\varphi_{2}(x_{0},E_{0})\}$ a basis of
$P(x_{0},E_{0})\C^{md}$ and we define:
$$\varphi_{1}(z,E)=P(z,E)\varphi_{1}(x_{0},E_{0})\quad;\quad \varphi_{2}(z,E)
=P(z,E)\varphi_{2}(x_{0},E_{0}).$$ Then, in a neighborhood of
$(x_{0},E_{0})$, $\{\varphi_{1}(z,E),\varphi_{2}(z,E)\}$ is an
analytic basis of $P(z,E)\C^{md}$. The matrix $M(z,E)$ of
$P(z,E)H(z,E,0)_{| P(z,E)\C^{md}}$ on the basis
$\{\varphi_{1}(z,E),\varphi_{2}(z,E)\}$ has analytic coefficients.
Besides, $\sigma(P(z,E)H(z,E,0)_{|
P(z,E)\C^{md}})=\{k_{i}(z,E,0),k_{j}(z,E,0)\}$. This implies that
$\det M(z,E)=k_{i}(z,E,0)k_{j}(z,E,0)$ and $\tr
M(z,E)=k_{i}(z,E,0)+k_{j}(z,E,0)$. We finish the proof with the
identity $(k_{i}-k_{j})^{2}=(k_{i}+k_{j})^{2}-4k_{i}k_{j}$. This
ends the proof of Lemma \ref{le:ana2}. \smallpagebreak We define:
$$\Lambda=\{(z,E)/\exists i\neq j/\quad (k_{i}-k_{j})(z,E,0)=0\}\subset\mathbb R\times\Delta.$$
According to \cite{Ka}, it suffices to prove that the functions
$\{k_{j}\}_{j=1,\dots, md}$ are analytic in a neighborhood of any
$(x_{0},E_{0})\in\Lambda$. Fix $(x_{0},E_{0})\in\Lambda$. There
exist $i$ and $j$ such that
$k_{i}(x_{0},E_{0},0)=k_{j}(x_{0},E_{0},0)$. For $l\neq j$ and
$l\neq i$, $k_{l}$ is analytic in a neighborhood of
$(x_{0},E_{0})$. By using Lemma \ref{le:ana2}, it suffices to
prove that $(z,E)\mapsto (k_{i}(z,E,0)-k_{j}(z,E,0))$ is analytic
at $(x_{0},E_{0})$.
 The function
$g(z,E)=(k_{i}-k_{j})^{2}(z,E,0)$ is analytic in a neighborhood
$V$ of $(x_{0},E_{0})$. Besides, since $k_{i}-k_{j}$ is real for
any $(x,E)$ in $V\cap\R^{2}$, we have:
$$g(x,E)\geq 0,\quad \forall (x,E)\in V\cap\R^{2}.$$
We write the Taylor expansion of $(z,E)\mapsto g(z,E)$. There
exist $(\alpha,\ \beta,\ \gamma)\in\R^{3}$ such that:
$$g(z,E)=[\alpha(z-x_{0})]^{2}+[\beta(z-x_{0})+\gamma(E-E_{0})]^{2}+o((z-x_{0})^{2}+(E-E_{0})^{2}).$$
We start with proving that there exist $(\tilde{\beta},\
\tilde{\gamma})\in\R^{2}$ such that
$$ g(z,E)=
[\tilde{\beta}(z-x_{0})+\tilde{\gamma}(E-E_{0})]^{2}+o((z-x_{0})^{2}+(E-E_{0})^{2}).$$
We know that, for any $E\in\Delta$, the function $z\mapsto
\sqrt{g(z,E)}$ is analytic. For $|E-E_{0}|$ and $|z-x_{0}|$ small
enough, we have that:
$$\sqrt{g(z,E)}=\sqrt{(\alpha^{2}+\beta^{2})\left(z-x_{0}+\frac{\gamma\beta(E-E_{0})}{\alpha^{2}+\beta^{2}}\right)^{2}+\frac{\gamma^{2}\alpha^{2}(E-E_{0})^{2}}{\alpha^{2}+\beta^{2}}+o((z-x_{0})^{2}+(E-E_{0})^{2})}.$$
The function $z\mapsto\sqrt{g(z,E)}$ can be analytically continued
in a neighborhood of $x_{0}$ only if $\gamma\alpha=0$. This proves
the announced result, with $(\tilde{\beta},\
\tilde{\gamma})=(\beta,\ \gamma)$ or $(\tilde{\beta},\
\tilde{\gamma})=(\sqrt{\alpha^{2}+\beta^{2}},0)$. \smallpagebreak
We notice that we have the following relations:
$$|\partial_{x}(k_{i}-k_{j})(x_{0},E_{0})|=|\tilde{\beta}|\neq 0.$$
$$|\partial_{E}(k_{i}-k_{j})(x_{0},E_{0})|=|\tilde{\gamma}|.$$
To end the proof, it remains to show that:
$$ g(z,E)=
[\tilde{\beta}(z-x_{0})+\tilde{\gamma}(E-E_{0})]^{2}+O(|\tilde{\beta}(z-x_{0})+\tilde{\gamma}(E-E_{0})|^{3}+|E-E_{0}|[\tilde{\beta}(z-x_{0})+\tilde{\gamma}(E-E_{0})]^{2}).$$
We change variables  for
$u=\tilde{\beta}(z-x_{0})+\tilde{\gamma}(E-E_{0})$ and
$e=(E-E_{0}).$ \smallpagebreak Since $\tilde{\beta}\neq 0$, this
map is bijective and we consider the function $\tilde{g}$:
$$\tilde{g}(u,e)=g\left(\frac{u-\tilde{\gamma}
e}{\tilde{\beta}}+x_{0},e+E_{0}\right).$$ We write the Taylor
expansion of $\tilde{g}$ near $(0,0)$:

$$\tilde{g}(u,e)=u^{2}+\sum_{l+q\geq 3}a_{lq}u^{l}e^{q}.$$
Since $\tilde{g}$ is real positive on a neighborhood of $(0,0)$,
we obtain that $a_{1q}=0$, for any $q\geq 2$. This implies:
$$\tilde{g}(u,e)=u^{2}\left(1+\sum_{l\geq 2,q\geq 1}a_{lq}u^{l-2}e^{q}\right)=u^{2}\left(1+O(|e|+|ue|)
\right).$$ \smallpagebreak Point $(3)$ is an immediate consequence
of $(2)$ and of Rouch\'e's Theorem. It ends the proof of Lemma
\ref{le:ana}.
\subsection{Proof of Lemma \ref{le:dev}}
A proof similar to the proof of Lemma \ref{le:ana2} shows that the
function $(z,E)\mapsto (k_{i}-k_{j})^{2}(z,E,\delta)$ is analytic
for $\delta$ small enough and that $(z, E,\delta)\mapsto
(k_{i}-k_{j})^{2}(z,E,\delta)$ is $C^{3}$. \smallpagebreak For
$\delta=0$, Lemma \ref{le:ana} implies that
$a(E)=|\partial_{z}(k_{i}-k_{j})(x_{0}(E),E,0)|$ is analytic in
$E\in\Delta$. For $\delta>0$, we define:
$$
r_{1}(z,E,\delta)=\frac{(k_{i}-k_{j})^{2}(z,E,\delta)-a^{2}(E)(z-x_{0}(E))^{2}}{2\delta}.
$$
The function $E\mapsto
\partial_{z}r_{1}(x_{0}(E),E,\delta)=c(E)+O(\delta)$ is analytic
and  $O(\delta)$ is uniform in $E$. Thus we can apply
Weierstrass's Theorem to get that the function $c$ satisfies
$c(E)=\lim_{\delta\rightarrow
0}\partial_{z}r_{1}(x_{0}(E),E,\delta)$ and is analytic in
$E\in\Delta$. We also define:
$$
r_{2}(z,E,\delta)=\frac{(k_{i}-k_{j})^{2}(z,E,\delta)-a^{2}(E)(z-x_{0}(E))^{2}-
2c(E)(z-x_{0}(E))\delta}{\delta^{2}}.$$ Similarly, the function
$b^{2}$ such that $b^{2}(E)=\lim_{\delta\rightarrow
0}r_{2}(x_{0}(E),E,\delta)$ is analytic.
\subsection{Proof of Lemma \ref{le:beh_a}}
\label{sec:techn3} We fix $\delta>0$ and drop it from the
notation. For $\rho>0$, we denote by
$V_{\rho}(\Delta)=\{\mathcal{E}\in\C\ ;\
\textrm{dist}(\mathcal{E},\Delta)<\rho\}$.\smallpagebreak By
perturbation theory, there exists $Y>0$ and $\rho>0$, depending on
$\delta$, such that $ (z,E)\mapsto k_{j}(z,E)$ is analytic on
$\rho_{Y}\times V_{\rho}(\Delta)$ and:
$$\inf\limits_{i\neq j,(z,E)\in \rho_{Y}\times V_{\rho}(\Delta) }(k_{i}(z,E)-k_{j}(z,E))>0.$$
Hence, for $a>0$ small enough, we can write the eigenprojector
$P_{j}(z,E)$ as:
$$P_{j}(z,E)=\frac{1}{2 i \pi }\int_{|\lambda-k_{j}(z,E)|=a}[H(z,E)-\lambda] ^{-1}d\lambda.$$
We recall the identity
$$[H(z,E)-\lambda] ^{-1}-[H(\infty,E)-\lambda] ^{-1}=[H(z,E)-\lambda] ^{-1}
[H(\infty,E)-H(z,E)][H(\infty,E)-\lambda] ^{-1} $$ and the fact
that $H(\cdot, E)$ and its derivatives with respect to $E$ satisfy
the analog of {\bf H2} uniformly in $E\in V_{\rho}(\Delta)$. This
implies that  for any $j\in \{1,\dots, md\}$, for any $l\in\mathbb
N$, and uniformly in $E\in V_{\rho}(\Delta)$:
\begin{equation} \label{esp} \sup\limits_{\re
z>0}|\re
z|^{2+\nu}\|\partial_E^l(P_{j}(z,E)-P_{j}(+\infty,E))\|+\sup\limits_{\re
z<0}|\re
z|^{2+\nu}\|\partial_E^l(P_{j}(z,E)-P_{j}(-\infty,E))\|<\infty.
\end{equation}
We consider the vector $\Phi_{j}(0,E)$ satisfying (\ref{eq:norm2})
for $z=0$ and which is analytic in $E$. Consider the identities
$$k_{j}(z,E)=\tr(P_{j}(z,E)H(z,E)),\quad
\Phi_{j}(z,E)=W(z,E)\Phi_{j}(0,E),$$ and  Cauchy formula:
$$\partial_{E}^{l}\partial_{x}^{p}[k_{j}(x,E)-k_{j}(\pm\infty,E)]=\frac{p!l!}{(2i\pi)^{2}}\oint_{|\mathcal{E}-E|=r}\oint_{|z-x|=r'}\frac{[k_{j}(z,\mathcal{E})-
k_{j}(\pm\infty,\mathcal{E})]}{(\mathcal{E}-E)^{l+1}(z-x)^{p+1}}dz
d\mathcal{E},$$ for $r$ and $r'$ small enough. Then,
\eqref{eq:deck} follows. To get \eqref{eq:decphi}, one also makes
use of the differential equation satisfied by $W(\cdot,E)$, of the
estimate (\ref{esp}) and one considers the first $d$ components of
$ \Phi_j(z,E)$ only.\smallpagebreak The explicit formula
\eqref{eq:mat_a} and the decay of $\partial_{x}k_{j}$ and of
$\partial_{x}\varphi_{j}$ yields formula \eqref{eq:deca}.\\
Finally, estimate (\ref{d3}) is a direct consequence of the
definitions (\ref{d1}) and (\ref{d2}) and of \eqref{eq:deck}.

\subsection{Proof of Lemma \ref{le:as}}
\label{sec:techn2} The proof of Lemma \ref{le:as} is virtually
identical to the one of Lemma 3.1 in \cite{JH:04}, once the
properties of the matrix $M(x,E,\varepsilon)$ have been
established. Therefore, we just give here the main steps of the
first part of the argument for the reader's convenience.
\smallpagebreak Because of (\ref{eq:deca}), we know that there
exists $C$ such that, uniformly in $E$ and $\varepsilon$,
 $$\int_{0}^{\infty}\|M(y,E,\varepsilon)\|dy<C.$$
Expressing the solutions of equation (\ref{eq:diff_c}) as Dyson
series:
\begin{eqnarray}\label{eq:dyson} && c(x,E,\varepsilon)\ =\ \sum_{n=0}^\infty
\int_0^x \int_0^{x_1}\cdots
 \int_0^{x_{n-1}}\\[3mm]
&& \quad \quad \quad \quad \quad \quad \times \
M(x_1,E,\varepsilon) M(x_2,E,\varepsilon)\cdots
M(x_n,E,\varepsilon)
 dx_1dx_2\cdots dx_n \ c(0,E,\varepsilon), \nonumber
\end{eqnarray}
we obtain the usual bound:
$$
\|\, c(x,E,\varepsilon)\,\|\ \leq\ e^{\int_0^\infty\,\|
M(y,E,\varepsilon)\|\,dy}\ \|\, c(0,E,\varepsilon)\,\|.
$$
Thus, we get from (\ref{eq:dyson}) that $c(x,E,\varepsilon)$ is
bounded as $x\rightarrow \pm\infty$. Next we show that
$\|c(x,E,\varepsilon)-c(y,E,\varepsilon)\|$ is arbitrarily small
for large $x$ and $y$, so that
$\lim_{x\rightarrow\infty}c(x,E,\varepsilon)=
c(\infty,E,\varepsilon)$ exists. It is enough to consider
$$
c(x,E,\varepsilon)\,-\, c(y,E,\varepsilon)\ =\ -\,\int_x^y\,
M(z,E,\varepsilon)\,c(z,E,\varepsilon)\,dz.
$$
The expression above with $y=\pm\infty$, and the properties of
$M$, $ c$, just proven yield the bound
$$c(x,E,\varepsilon)-c(\pm\infty,E,\varepsilon)=O(<x>^{-(1+\nu)}).$$
Finally,  if $\|c(\pm\infty,E,\varepsilon)\|$ is further assumed
to be uniformly bounded in $E\in\Delta$ and
$\varepsilon\rightarrow 0$, it is enough to consider the initial
conditions (\ref{inco}), by linearity. Then, by integration by
parts, see (\ref{eq:intpp}) and (\ref{eq:ifdis}), we get,
$$
c(x,E,\varepsilon)\ =\ O(1),
$$
uniformly in $E\in\Delta$, $\varepsilon\rightarrow 0$ and
$x\in\mathbb R$. Hence, all bounds above are uniform in
$E\in\Delta$ and $\varepsilon\rightarrow 0$. \smallpagebreak
Getting similar bounds on the derivatives  of $c$ with respect to
$E$ which are uniform in $\varepsilon$ and $E\in\Delta$ requires a
little more work. The argument is identical to that used in
\cite{JH:04}.  We resort again to integration by parts in
(\ref{eq:diff_d}) with $x_0=-\infty$, differentiate with respect
to $E$, and make use of Gronwall Lemma to get bounds. We do not
give the details and refer to Section $7$ of \cite{JH:04}.
\subsection{Proof of Lemma \ref{le:large_t}}
\label{sec:large_t} Again, we mimic the proof of Lemma 4.1 in
\cite{JH:04}. It suffices to rewrite:
$$e^{-i(\int_{0}^{x}k_j(y,E)dy+tE)/\epsilon} \ =\
i\epsilon\ \frac{\frac{\partial}{\partial
E}e^{-i(\int_{0}^{x}k_j(y,E)dy+tE)/\epsilon}}
{\left(t+\int_{0}^{x} \frac{\partial}{\partial
E}k_j(y,E)dy\right)}.$$ We compute:
\begin{eqnarray}\phi_{j}(x,t,\varepsilon)&=&i\varepsilon\left[\frac{c_{j}(x,E,\varepsilon)
Q(E,\varepsilon)\varphi_{j}(x,E,\varepsilon)}{t+\int_{0}^{x}\partial_{E}k_j(y,E)dy}
e^{-i(\int_{0}^{x}k_j(y,E)dy+tE)/\varepsilon}\right]_{\partial\Delta} \nonumber \\
&-&i\varepsilon\int_{\Delta}\frac{\partial_{E}[c_{j}(x,E,\varepsilon)Q(E,\varepsilon)
\varphi_{j}(x,E,\varepsilon)]}{t+\int_{0}^{x}\partial_{E}k_j(y,E)dy}
e^{-i(\int_{0}^{x}k_j(y,E)dy+tE)/\varepsilon}dE\nonumber\\
&+&i\varepsilon\int_{\Delta}\frac{c_{j}(x,E,\varepsilon)Q(E,\varepsilon)
\varphi_{j}(x,E,\varepsilon)e^{-i(\int_{0}^{x}k_j(y,E)dy+tE)/\varepsilon}}
{[t+\int_{0}^{x}\partial_{E}k_j(y,E)dy]^{2}}\int_{0}^{x}\partial_{E}^{2}k_j(y,E)dy
dE.
\end{eqnarray}
By  Lemma \ref{le:beh_a} we have for $x$ large enough:
$$ t+\int_{0}^{x}\partial_{E}k_j(y,E)dy=x\left[\partial_{E}k_j(\pm\infty,E)+\frac{t}{x}+
O\left(\frac{1}{x}\right)\right].$$ Under the restrictions put on
$x/t$, we obtain that for a constant $C_\varepsilon$ independent
of $t$:
$$\|\phi_{j}(x,t,\varepsilon)\|\leq\frac{C_\varepsilon}{|x|}.$$
Now, for any $l\in\{0,\cdots,m-1\}$, we have:
$$\phi_{j}^{[l]}(x,t,\varepsilon)=\int_{\Delta}c_{j}(x,E,\varepsilon)k_{j}^{l}(x,E)\varphi_{j}(x,E)e^{\frac{-i(\int_{0}^{x}k_{j}(y,E)dy+Et)}{\varepsilon}}Q(E,\varepsilon)\ dE,$$
so that a similar computation gives the result.

The last estimate makes use of the fact that under the given
conditions stated on the signs of $x$ and $t$, and for $|x|$ large
enough, uniformly in time and energy,
\begin{equation}
| t+\int_{0}^{x}\partial_{E}k_j(y,E)dy|\geq
|t+x\partial_{E}k_j(\pm\infty,E)|/2\geq (|t|+|x|)c,
\end{equation}
for some  $c>0$. The result follows from the elementary inequality
$(x+y)\geq x^\beta y^{1-\beta}$, for any $x,y\geq 0$ and any
$\beta\in (0,1)$, and the arguments used above.

\subsection{Proof of Proposition \ref{pro:asl_2}}
\label{sec:asl_2} We adapt the proof of Proposition 4.1 in
\cite{JH:04} and we only give the main steps.

i) We rewrite:
\begin{eqnarray}
\label{eq:diff_phi2}
&&\phi_{j}^{[l]}(x,t,\varepsilon)-\phi_{j}^{[l]}(x,t,\varepsilon,\pm\infty)=\\
&&\int_{\Delta}[k_{j}^{l}(x,E)\varphi_{j}(x,E)-k_{j}^{l}(\pm\infty,E)\varphi_{j}(\pm\infty,E)]e^{\frac{-i\int_{0}^{x}k_{j}(y,E)dy}{\varepsilon}}c_{j}(x,E,\varepsilon)
Q(E,\varepsilon)e^{-itE/\varepsilon}\, dE+\nonumber\\
&&\int_{\Delta}k_{j}^{l}(\pm\infty,E)\varphi_{j}(\pm\infty,E)e^{\frac{-i(k_{j}(\pm\infty,E)x+\omega_{j}(\pm\infty,E))}{\varepsilon}}[e^{-ir_{j}^{\pm}(x,E)/\varepsilon}-1]c_{j}(x,E,\varepsilon)Q(E,\varepsilon)e^{-itE/\varepsilon}\, dE +\nonumber\\
&&\int_{\Delta}k_{j}^{l}(\pm\infty,E)\varphi_{j}(\pm\infty,E)e^{\frac{-i(k_{j}(\pm\infty,E)x+\omega_{j}(\pm\infty,E))}{\varepsilon}}[c_{j}(x,E,\varepsilon)-c_{j}(\pm\infty,E,\varepsilon)]Q(E,\varepsilon)e^{-itE/\varepsilon}\,
dE.\nonumber
\end{eqnarray}

ii) To obtain a factor $1/t$, we integrate by parts. For any
regular function $f$, we have:
\begin{equation}\label{eq:ipp}
\int_{\Delta} f(x,E,\varepsilon)e^{-itE/\varepsilon}dE
=\left[\frac{i \varepsilon}{t}\ f(x,E,\varepsilon)\
e^{-itE/\varepsilon}\,\right]_{\partial\Delta}
-\frac{i\varepsilon}{t}\ \int_{\Delta}\
\partial_{E}f(x,E,\varepsilon)\ e^{-itE/\varepsilon}dE.
\end{equation}

iii) We apply equation (\ref{eq:ipp}) for the three terms in
(\ref{eq:diff_phi2}).

iv) The estimates of Lemma \ref{le:beh_a} and Lemma \ref{le:as}
prove that there exists $C_\varepsilon>0$ such that:
$$\sup\limits_{x\rightarrow\pm\infty}|x|^{\nu}\|[\phi_{j}^{[l]}(x,t,\varepsilon)-
\phi_{j}^{[l]}(x,t,\varepsilon,\pm\infty)]\|<\frac{C_\varepsilon}{|t|}.$$

v) The identity
$\phi_{j}^{[l]}(x,t,\varepsilon)=\omega(x)\phi_{j}^{[l]}(x,t,\varepsilon)
+(1-\omega(x))\phi_{j}^{[l]}(x,t,\varepsilon)$ and support
considerations on the definition of
$\phi_{j}^{[l]}(x,t,\varepsilon, a)$ yield the first result.

vi) The second estimate follows from (\ref{eq:ipp}) applied to the
definition of $\phi_{j}^{[l]}(x,t,\varepsilon)$.


\subsection{Proof of Theorem \ref{th:astrans}}
Taking into account the results of Section \ref{sec:t_depdt}, we
are left with the  computation of the following integral:
$$T(\varepsilon,x,t)=\int_{\Delta}\tilde{P}(E,\varepsilon)e^{-\alpha(E)/\varepsilon}e^{-i\kappa(E)/\varepsilon}e^{-itE/\varepsilon}e^{-ixk_{n}(+\infty,E)/\varepsilon}\varphi_{n}(+\infty,E)dE,$$
with
$\tilde{P}(E,\varepsilon)={P}(E,\varepsilon)e^{-i\theta_j(\zeta,E
)}(1+O_E(\varepsilon))$, where $O_E(\varepsilon)$ is defined in
(\ref{eq:choice2}). We already remarked that
$\tilde{P}(E,\varepsilon)$ satisfies {\bf C3} as well. Again, wa
adapt the arguments of \cite{JH:04}.\smallpagebreak
 In terms of the variable
$k=k_{n}(+\infty,E)$, we rewrite $T$ as:
$$\int_{k_{n}(\Delta,+\infty)}\tilde{P}(E_{n}^+(k),\varepsilon)e^{-\alpha(E_{n}^+(k))/\varepsilon}e^{-i\kappa(E_{n}^+(k))/\varepsilon}e^{-itE_{n}^+(k)/\varepsilon}e^{-ixk/\varepsilon}\varphi_{n}(+\infty,E_{n}^+(k))(\partial_{k}E_{n}^+)(k)dk.$$
We can see $T$ as the rescaled Fourier transform (see
(\ref{eq:fourier})) of the function:
$$S(\varepsilon,k,t)=\sqrt{2\pi\varepsilon}\tilde{P}(E_{n}^+(k),\varepsilon)e^{-\alpha(E_{n}(+\infty,k))/\varepsilon}e^{-i\kappa(E_{n}^+(k))/\varepsilon}e^{-itE_{n}^+(k)/\varepsilon} \varphi_{n}(+\infty,E_{n}^+(k))(\partial_{k}E_{n}^+)(k).$$

We follow the analysis done in \cite{JH:04} and expand $S$ around
$k^{*}$
$$
\alpha(E_{n}^+(k))=\alpha(E^{*})+[\partial_{k}E_{n}^+(k^{*})]^{2}\alpha"(E^{*})\frac{(k-k^{*})^{2}}{2}+O((k-k^{*})^{3}),$$
where $\alpha"(E^{*})>0.$ We define:
$$\tilde{\kappa}(k)=\kappa(E_{n}^+(k)),$$
$$\tilde{\kappa}(k)=\tilde{\kappa}(k^{*})+\tilde{\kappa}'(k^{*})(k-k^{*})+\tilde{\kappa}"(k^{*})\frac{(k-k^{*})^{2}}{2}+O((k-k^{*})^{3}).$$
We also have:
$$\tilde{P}(E_{n}^+(k),\varepsilon)=P(E_{n}^+(k^{*}),\varepsilon)
e^{-i\theta_j(\zeta,E_{n}^+(k^{*})}+O((k-k^{*}))
+O(\varepsilon).$$
$$(\partial_{k}E_{n}^+)(k)\varphi_{n}(+\infty,E_{n}^+(k))=(\partial_{k}E_{n}^+)(k^{*})\varphi_{n}(+\infty,E_{n}^+(k^{*}))+O((k-k^{*})).$$
We can then follow the computations of \cite{JH:04}. We set
$\mu(\varepsilon)=\varepsilon^s$, with $1/3< s < 1/2$, so that
$\mu(\varepsilon)^3/\varepsilon << 1 <<
\mu(\varepsilon)/\sqrt{\varepsilon}$, as $\varepsilon\rightarrow
0.$ Using Parseval formula and thanks to the properties of
$\alpha$, the restriction of the integration set to the interval
$[k^*- \mu(\varepsilon), k^*+ \mu(\varepsilon)]$ causes an error,
in the $L^2$ sense, of order
$e^{-\alpha(E^*)/\varepsilon}\varepsilon^\infty$. On that set, the
overall error $\mbox{err} (\varepsilon)$ stemming from the
expansion around $k^*$ is $\mbox{err}
(\varepsilon)=O(\varepsilon+\varepsilon^s+\varepsilon^{3s-1})=
O(\varepsilon^{3s-1})$, with our values of $s$. The $L^2$ norm of
the error term can thus be estimated by Parseval formula to yield
$$
\left\|\int_{[k^*- \mu(\varepsilon), k^*+ \mu(\varepsilon)]}
\mbox{err}
(\varepsilon)e^{-\alpha(E^*)/\varepsilon}e^{-\Lambda(k)/\varepsilon}e^{-itE_{n}^+(k)/\varepsilon}e^{-ixk/\varepsilon}\
dk\right\| =O\left(\mbox{err}
(\varepsilon)e^{-\alpha(E^*)/\varepsilon}\varepsilon^{3/4}\right),$$
uniformly in $t$. The interval of integration can then be restored
to its initial value $k_n(+\infty, \Delta)$ at the expense of
another error of order $e^{-\alpha(E^*)/\varepsilon}
\varepsilon^\infty.$ Therefore, we obtain that there exists
$3/4<p<5/4$ so that in $L^{2}$-norm and with an error term that is
uniform in $t$:
\begin{eqnarray}
T(\varepsilon,x,t)&=&e^{-\alpha(E^{*})/\varepsilon}P(E^{*},\varepsilon)
e^{-i\theta_j(\zeta,E^*)}e^{-i\tilde{\kappa}(E^{*})/\varepsilon}
\varphi_{n}(+\infty,E^{*})\partial_{k}E_{n}^+(k^{*})\times\\
&\times&\int_{k_{n}(+\infty,\Delta)}e^{-\Lambda(k)/\varepsilon}e^{-itE_{n}^+(k)/\varepsilon}e^{-ikx/\varepsilon}dk
+O(e^{-\alpha(E^{*})/\varepsilon}\varepsilon^{p}).\nonumber
\end{eqnarray}
\subsection{Proof of Lemma \ref{le:quad}}
To prove Lemma \ref{le:quad}, it suffices to compute:
$$\int_{k_{n}(+\infty,\Delta)}e^{-\Lambda(k)/\varepsilon}e^{-itE_{n}^{+}(k)/\varepsilon}e^{-ikx/\varepsilon}dk.$$
According to (\ref{eq:quad}), we have:
$$E_{n}^{+}(k)=E_{n}^{+}(k^{*})+\partial_{k}E_{n}^{+}(k^{*})(k-k^{*})+\frac{\partial^{2}_{k}E_{n}^{+}(k^{*})}{2}(k-k^{*})^{2}.$$
With the same arguments as in \cite{JH:04}, we can expand the
interval of integration to the whole line at the expense of an
error that is uniform in $t$, and in the $L^2$ sense:
$$\int_{k_{n}(+\infty,\Delta)}e^{-\Lambda(k)/\varepsilon}e^{-itE_{n}^{+}(k)/\varepsilon}e^{-ikx/\varepsilon}dk=\int_{-\infty}^{+\infty}e^{-\Lambda(k)/\varepsilon}e^{-itE_{n}^{+}(k)/\varepsilon}e^{-ikx/\varepsilon}dk+O(\varepsilon^{\infty}).$$
  The rest of the proof follows from the formula:
$$
\int_{-\infty}^\infty
\,e^{-(M(k-k^*)^2/2+iN(k-k^*))/\varepsilon}\, dk\ =\
\sqrt{\frac{\varepsilon 2\pi}{M}}\ e^{-\frac{N^2}{2\varepsilon
M}},
$$
with
$$M=\lambda_{2}+i\partial_{k}^{2}E_{n}^{+}(k^{*})t,$$
$$N=\lambda_{1}+\partial_{k}E_{n}^{+}(k^{*})t+x.$$
\subsection{Proof of Proposition \ref{stph}}
We will make use of the following Lemma whose proof we omit.
\begin{Le}\label{71} Let $f$ and $u_\varepsilon$ be
$C_0^\infty(\mathbb R, \mathbb R)$. Provided $f'(k)\neq 0$ on
$\mbox{\em supp} \ u_\varepsilon$, we have for any $n\in\mathbb N$
$$
\int_{\mathbb R} e^{-if(k)/\varepsilon}u_\varepsilon(k)\ dk=
(-i\varepsilon)^n\int_{\mathbb R} e^{-if(k)/\varepsilon}
\left(\left(\cdots\left(\left(\frac{u_\varepsilon(k)}{f'(k)}\right)'\frac{1}{f'(k)}\right)'
\cdots \right)'\frac{1}{f'(k)}\right)'\, dk \ ,
$$
where there are $n$ factors $1/f'(k)$. Moreover, setting $g=1/f'$,
there exist constants $c(j,r_1, \cdots, r_n)\in\mathbb R$ such
that
$$
\left(\left(\cdots\left(\left({u_\varepsilon}{g}\right)'g\right)'
\cdots \right)'g\right)'=\sum_{j=0}^n\sum_{{r_1, \cdots, r_n \atop
\sum_{i=1}^nr_i=j}} c(j,r_1, \cdots, r_n)
u_\varepsilon^{(n-j)}g^{(r_1)}g^{(r_2)}\cdots g^{(r_n)},
$$
where $h^{(k)}$ denotes the $k^{\mbox{th}}$ derivative of $h$.
\end{Le}
In our case, $f'(k)=x+\partial_kE^\sigma_l(k)t$, so that explicit
computations show the existence of smooth functions on
$S_\varepsilon$, the support of
$u_\varepsilon=e^{-\Lambda(k)/\varepsilon}\eta_\varepsilon$,
$k\mapsto c_{j,n}(k)$ which are independent of $x$ and $t$ and
such that
\begin{equation}\label{eq:derf}
\left(\frac{1}{f'(k)}\right)^{(n)}=\sum_{j=0}^nc_{j,n}(k)\frac{t^j}
{(x+\partial_kE^\sigma_l(k)t)^{j+1}}.
\end{equation}
Then, we get from the above
\begin{Le} Assume $\partial_kE^\sigma_l(\tilde{k})\neq 0$ and consider
$(x,t)$ such that $\inf_{k\in
S_\varepsilon}|x+\partial_kE^\sigma_l(k)t|>0$. For any
$n\in\mathbb N$, there exists a positive constant $C(n)$, uniform
in $(x, t)$ and $\varepsilon$ such that
\begin{equation}\label{rhs}
\left| \int_{\mathbb R}\
\,e^{-i(kx+tE_l^\sigma(k))/\varepsilon}e^{-\Lambda(k)/\varepsilon}\eta_\varepsilon(k)\,dk\right|
\leq {C(n)\varepsilon^\tau} \sum_{r=0}^n\frac{(\varepsilon |t|)^r}
{\left(\inf_{k\in
S_\varepsilon}|x+\partial_kE^\sigma_l(k)t|\right)^{r+n}},
\end{equation}
for all $\varepsilon<\varepsilon_0$, where $\varepsilon_0$ is
uniform in $n$ and $(x,t)$.
\end{Le}
{\bf Remark:} The Lemma actually holds for any value of
$\varepsilon$ if one is not interested in the behavior in that
parameter. \demo One first notes that by scaling and by our choice
of $\tau$ the successive derivatives of $u_\varepsilon=
e^{-\Lambda(k)/\varepsilon}\eta_\varepsilon$ satisfy
$\sup_{k\in\mathbb R}u_\varepsilon^{(j)}(k)=O(1/\varepsilon^{j})$,
if $\varepsilon$ is small enough. Then, formula (\ref{eq:derf})
and the restriction $\sum_{i=1}^n r_i=j$ in the second point of
Lemma \ref{71} give the possible numbers of factors
$\frac{t^j}{(x+\partial_kE^\sigma_l(k)t)^{j+1}}$ in the integrand.
Finally, the fact that all functions of $k$ involved are smooth,
together with $|S_\varepsilon|=\varepsilon^\tau$ yield the result.

To end the proof of the Proposition, we note that on the
complement of $C_t(\varepsilon)$, the right hand side of
(\ref{rhs}) is bounded above by
$$
\frac{C(n)\varepsilon^\tau \varepsilon |t|^{1-\alpha}}{(\inf_{k\in
S_\varepsilon} |x+\partial_kE^\sigma_l(k)t|)}
\left(\frac{\varepsilon}{|t|^{2\alpha -1}}\right)^{n-1},
$$ using the fact that $\varepsilon |t|^{1-\alpha}>1$.
Explicit computation of the $L^2$ norm of $(\inf_{k\in
S_\varepsilon} |x+\partial_kE^\sigma_l(k)t|)$ on the complement of
$C_t(\varepsilon)$ yields the result.

\end{document}